\documentclass[11pt,a4,epsf,fleqn]{article}
\usepackage{verbatim}
\usepackage{eurosym,rotating,float}
\usepackage{wrapfig, framed, caption}
\usepackage{caption}
\usepackage{todonotes}
\usepackage{soul}
\usepackage{mathcomp}
\usepackage{textcomp}
\usepackage[para,online,flushleft]{threeparttable}
\usepackage{longtable}
\usepackage{bm}
\usepackage[sumlimits,fleqn]{amsmath}
\usepackage{array}
\newcolumntype{K}[1]{>{\centering\arraybackslash}m{#1}}
\usepackage[sort&compress]{natbib}
\usepackage[final]{pdfpages}
\usepackage{booktabs}
\usepackage{arydshln}
\usepackage[bookmarksopen,bookmarksnumbered,colorlinks,linkcolor={blue},citecolor={blue}]{hyperref}
\usepackage{cleveref}[2012/02/15]
\crefformat{footnote}{#2\footnotemark[#1]#3}
\usepackage{graphicx}
\usepackage{color}
\newcolumntype{P}[1]{>{\centering\arraybackslash}p{#1}}
\newcolumntype{L}{>{\centering\arraybackslash}m{3cm}}
\newcolumntype{M}[1]{>{\centering\arraybackslash}m{#1}}
\usepackage{multirow}
\usepackage{lscape}
\usepackage{pdflscape}
\usepackage[T1]{fontenc}
\usepackage{pdflscape}
\usepackage{adjustbox}
\usepackage{mathtools}
\usepackage{enumitem}
\setlist[description]{leftmargin=\parindent,labelindent=\parindent-0.2cm}
\setlist{noitemsep}
\DeclareMathSymbol{\mathdblquotechar}{\mathalpha}{letters}{`"}

\newcommand{\mathdblquote}{\mathtt{\mathdblquotechar}}
\begingroup\lccode`~=`"\lowercase{\endgroup
  \let~\mathdblquote
}
\AtBeginDocument{\mathcode`"="8000 }
\usepackage{amsmath}

 \usepackage{listings}
\begin{document}
\title{\vspace{-3.5cm}Statistical significance in choice modelling:\\ computation, usage and reporting}

\author{Stephane Hess\thanks{s.hess@leeds.ac.uk; Institute for Transport Studies and Choice Modelling Centre, University of Leeds} \and Andrew Daly\thanks{andrew@alogit.com; Institute for Transport Studies and Choice Modelling Centre, University of Leeds} \and Michiel Bliemer\thanks{michiel.bliemer@sydney.edu.au; Institute of Transport and Logistics Studies, The University of Sydney} \and Angelo Guevara\thanks{crguevar@gmail.com; Department of Civil Engineering, Universidad de Chile} \and Ricardo Daziano\thanks{daziano@cornell.edu; School of Civil and Environmental Engineering, Cornell University} \and Thijs Dekker\thanks{t.dekker@leeds.ac.uk; Institute for Transport Studies and Choice Modelling Centre, University of Leeds} }

\maketitle

\vspace{-1cm}
\begin{abstract}
This paper offers a commentary on the use of notions of statistical significance in choice modelling. We review the reasons for uncertainty in parameter estimates, provide a precise discussion on the computation of measures of uncertainty and confidence intervals, and discuss the use of statistical tests. We argue that, as in many other areas of science, there is an over-reliance on 95\% confidence levels, and misunderstandings of the meaning of significance. We also observe a lack of precision in the reporting of measures of uncertainty in many studies, especially when using $p$-values and even more so with \emph{star} measures. The paper also stresses the importance of considering behavioural or policy significance in addition to statistical significance. Finally, we stress a number of points that are specific to choice modelling and which require special attention, notably in relation to derived measures such as willingness-to-pay, the treatment of random heterogeneity, and the use of repeated choice data.\\

\emph{Keywords:} choice modelling; standard errors; $t$-ratios; $p$-values; confidence intervals
\end{abstract}

\section{Introduction}

\noindent Empirical work focusses on producing metrics of interest, with a particular emphasis on investigating the existence of specific effects or phenomena. In this context, analysts often formulate hypotheses, and then run tests to establish whether an effect exists. When using mathematical structures (models) to represent the data, tests also look at whether one model explains the data better than another. With finite and limited data, any empirical findings have uncertainty attached to them, and the recognition of this leads analysts to attach measures of confidence to their findings, under the broad umbrella of \emph{``statistical significance''.}

While ubiquitous across different methodological disciplines and fields of application, there is also growing criticism of the over-reliance on statistical significance, and its misuse, in several branches of science. This is exemplified in the wide ranging discussions by \citet{Ziliak}, or more recent papers calling for care in interpretation \citep[see e.g.][]{ASA}, the use of much stricter tests \citep{Benjamin_et_al}, or even a complete abandonment of the notion of statistical significance \citep{Amrhein_et_al,Wasserstein_et_al}. 

A key component of this criticisms relates to excessive attention to the question whether an effect exists (i.e. is it different from zero) rather than whether an effect matters (i.e. how large it is). Similarly, there is misunderstanding about what statistical tests tell us, as highlighted in the discussions by \citet{Ziliak} on the ``fallacy of the transposed conditional'', with a clear distinction between the probability of the evidence given a hypothesis and the probability of the hypothesis given the evidence. 

Against this background, the timing seems apt to offer a commentary on this issue within the field of choice modelling. A number of specific considerations arise in our field, making a separate discussion appropriate. We list six of these reasons in what is a non-exhaustive list:

\begin{enumerate}
    \item Choice modelling relies on the results of a \emph{model} for which parameters are \emph{estimated}, meaning that measures of uncertainty cannot simply be calculated on the basis of the observed data. Instead, they rely on statistical theory, in particular in relation to the properties of maximum likelihood estimates (MLE), or the assumed data generating process and selected priors in Bayesian analysis. That is, also in Bayesian analysis, uncertainty is present, and even more specific. 
    \item Estimates from choice models have a different interpretation from for example estimates of a linear regression model, and this means that analysts rely on measures calculated on the basis of parameter transformations for actual insights, such as marginal rates of substitution. Other insights are obtained using model application, such as forecasting, especially if an analyst wants to establish the size of an effect beyond its existence per se. These outputs again need measures of uncertainty.
    \item Applied choice modelling relies on the recognition and flexible treatment of extensive levels of heterogeneity in underlying preferences, leading to a requirement for a distinction between parameter uncertainty and distribution of preferences in a sample population.
    \item There is a strong reliance on panel (or repeated choice) data, requiring either an explicit recognition of correlation across choices for the same person, or ex-post correction approaches to measures of uncertainty computed based on statistical theory developed for cross-sectional data.
    \item A key focus in choice modelling relates to tests of structural assumptions of the underlying models, with extensive model comparison work where a number of tests have been developed for choice models in particular.
    \item Finally, in addition to these field-specific considerations, it should also be recognised that the rapid expansion of choice modelling over the last couple of decades has opened the method up to analysts from a diverse set of backgrounds, including those with limited econometric training. The key underlying statistical concepts remain poorly understood by some, and statistical test and measures of uncertainty are used and reported inappropriately by many.
\end{enumerate}

\noindent An important question arises before proceeding, namely \emph{why write a paper on significance measures and testing in the face of growing concerns about the usefulness of such tests?} For example, \citet{Imbens_2021} argues that \emph{``...researchers are interested in a point estimate and the degree of uncertainty associated with that point estimate as the precursor to making a decision or recommendation to implement a new policy. In such cases, the absence or presence of statistical significance (in the sense of being able to reject the null hypothesis of zero effect at conventional levels) is not relevant''}. Even more strongly, there are strong arguments against the usefulness of a hypothesis of a zero effect. \citet{abadie2020statistical} state that \emph{``in economics ... there are rarely reasons to put substantial prior probability on a point null''} while \citet{gelman2020regression} state that \emph{``... we do not generally think null hypotheses can be true: in social science and public health, just about every treatment one might consider will have some effect, and no comparison or regression coefficient of interest will be exactly zero.''}

First, the usefulness of measures of uncertainty extends beyond statistical tests to the computation of confidence intervals, and for this, the background on computation is crucial. Second, we are of the opinion that notwithstanding the validity of some of these concerns, choice modellers will continue relying on statistical tests in this way, and our note at least serves to improve the way in which they do this, for example using one-sided tests in many cases.

Our paper addresses the topic along a number of different dimensions. To provide a stronger foundation for understanding the notion of uncertainty, Section \ref{sec:uncertainty} discusses the reasons for measures of uncertainty and their calculation. The meaning of uncertainty is then explained further in the context of confidence intervals in Section \ref{sec:confidence_intervals}. The central part of the paper, namely Section \ref{sec:significance_tests}, focusses on statistical tests. We discuss the formation and testing of hypothesis, separately for individual results and for model selection, before turning to the reporting of uncertainty and test outcomes. In our discussions, we question the strict adherence to the 95\% level of confidence, highlight misconceptions as to what this implies, and highlight a lack of precision in some reporting practices. We also stress the need for more precise language in describing the outcome of significance tests. In this context, we also caution analysts to pay attention to the size of effects rather than the existence of effects alone, thus also considering the importance of a finding from a behavioural or policy perspective. While the points raised in this paper will be obvious to many choice modellers with a strong econometric background, this seems to not be the case in the more applied community. This conclusion is supported by the recent discussions in \citet{Parady_2023} who note a lack of rigour in the reporting of results and their statistical properties. To support the theoretical discussions, Section \ref{sec:application} presents an empirical example, before Section \ref{sec:conclusions} offers some conclusions.

\section{Measures of uncertainty}\label{sec:uncertainty}

\subsection{Sampling error and parameter uncertainty}

Our specific interest relates to model parameters (and later transformations thereof). The frequentist (or classical) view holds that, for each single parameter $\beta_k$, there exists, subject to the chosen model specification, a unique true (or population) value $\beta_k^*$, but that our data is incomplete (it is a sample) and we thus have sampling error in our estimate $\hat{\beta}_k$ for parameter $\beta_k$. An estimate without uncertainty would thus only be obtained if the entire population was sampled. 

The key concept in classical statistics corresponds to the ``sampling distribution'' of $\hat{\beta}_k$, the estimate of $\beta_k$. Specifically, $\hat{\beta}_k$ is a random variable, because, if a different sample were taken, a different point estimate $\hat{\beta}_k$ would be obtained. If an analyst were to repeat the sampling process many times (say $S$), giving estimates $\hat{\beta}_k^{(1)},\hdots,\hat{\beta}_k^{(S)}$, a full ``sampling distribution'' of $\hat{\beta}_k$ would be revealed as $S$ grows, with its respective mean $\mu_{\hat{\beta}_k}$ and standard deviation $\sigma_{\hat{\beta}_k}$. It could be shown, and should be intuitively clear that, with larger sample sizes $\left(N\right)$, the variance of the sampling distribution would decrease, as a larger share of the population would be included in each sample.

In practice, an analyst would of course not repeat the sampling process to build the sampling distribution, but would estimate the model on the entire set of collected data, and would also obtain, from the same sample, estimators $\hat\mu_{\beta_k}$ and $\hat{\sigma}_{\beta_k}$ of the parameters of the sampling distribution. We next turn to how this is derived in the context of maximum likelihood estimation.

\subsection{From maximum likelihood estimation to the asymptotic covariance matrix}

While a brief Bayesian perspective is offered in the concluding points, our discussion primarily centres on classical estimation and inference. Some basic understanding of maximum likelihood estimation is thus helpful for the remainder of this paper. 

Let us assume that we have a sample of $N$ people\footnote{Here, $N$ relates to the sample actually used in estimation, i.e. excluding any hold out data set aside for validation.}. Let us further denote the chosen alternative for person $n$ as $Y_{n}$. The analyst specifies a model that uses a vector of model parameters $\beta$ (of length $K$), with $\beta=\left<\beta_1,\hdots,\beta_K\right>$. The log-likelihood function for this model is then given by:
\begin{equation}\label{eq:LL}
LL_N\left(\beta\right)=\sum_{n=1}^N \log L_{n}\left(Y_n\mid\beta\right),
\end{equation}
where $L_{n}\left(Y_n\mid\beta\right)$ is the likelihood of the observed choice for person $n$, where the subscript $N$ in $LL_N(\beta)$ denotes the sample size. The specific functional form for $L_{n}\left(Y_n\mid\beta\right)$ will vary across models. 

The use of maximum likelihood (ML) estimation of a discrete choice model (DCM) on a given sample yields the maximum likelihood estimator (MLE) $\hat{\beta}$ as:
\begin{equation}\label{eq:mle}
\hat{\beta}=\underset{\beta}{\arg\max} \,LL_N\left(\beta\right).	
\end{equation}
For DCMs, we can only make asymptotic statements about the sampling distribution, that is, statements that relate to a case when the sample size $N$ goes to infinity. In addition, it should already be noted that this theory also relies on the conditions that 1) the model is correctly specified, and 2) that the model is identified, i.e.\ not overspecified, thus for example requiring appropriate normalisations.

Under these conditions, as sample size $N$ increases, the MLE for our 
vector $\beta$, i.e. $\hat{\beta}$, converges to a normal distribution\footnote{We refer the reader to \citet[][chapter 5]{alma990024527500403126}.} around the vector of true population values $\beta^*$, with:
\begin{equation}\label{eq:asymptotic}
\sqrt{N}\left(\hat{\beta}-\beta^*\right)\rightarrow \mathcal{N}\left(0,A^{-1} I A^{-1}\right),
\end{equation}
such that $\hat\beta$ has asymptotic variance matrix $\Omega = \frac{1}{N}A^{-1} I A^{-1}$,  
where:
\begin{align}
A&=E\big(H\big), \quad \text{where} \quad
H = \frac{\partial^2 LL_N(\beta^*)}{\partial\beta\text{ }\partial\beta'}, \qquad \text{ and}\\
I&=E\big(S S'\big), \quad \text{where} \quad
S = \frac{\partial LL_N(\beta^*)}{\partial\beta},
\end{align}
where $H$ is the matrix of second derivatives of the log-likelihood function, also referred to as the {\it Hessian}, $S$ is the vector of first derivatives of the log-likelihood function, also referred to as the {\it score}, both evaluated in the true values $\beta^*$. The expectation is over choice observations $Y$. Matrix $I$ is the variance of the score and is called the Fisher information matrix, which measures the amount of information that the choice observations $Y$ carry about the unknown parameters. 

Under mild regularity conditions \citep[see e.g.][Lemma 5.3]{LehmannCasella1998}, it holds that $I = -A$, such that $\Omega = \frac{1}{N}A^{-1} I A^{-1}$ simplifies to $\Omega = \frac{1}{N}I^{-1}$ or $\Omega = -\frac{1}{N} A^{-1}$. 
To determine a sample estimate of the asymptotic variance matrix evaluated at $\hat\beta$, denoted by  
$\hat{\Omega}$, one can compute $\hat{A}$ and $\hat{I}$,
\begin{align}
\hat{A}&=\frac{1}{N} \hat{H}, \quad \text{where} \quad
\hat{H} = \sum_{n=1}^N\frac{\partial^2 \log L_n(\hat{\beta})}{\partial\beta\text{ }\partial\beta'}, 
 \qquad \text{and}\\
\hat{I}&=\frac{1}{N}\hat{O}, \quad \text{where} \quad 
\hat{O} = \sum_{n=1}^N\frac{\partial \log L_n(\hat{\beta})}{\partial\beta}\frac{\partial \log L_n(\hat{\beta})}{\partial\beta'}\label{eq:BHHH},
\end{align}
where $\hat{H}$ is the Hessian and $\hat{O}$ is the sum of outer products of partial derivatives, both evaluated in the estimates $\hat\beta$. We can compute three different estimates of the variance matrix as shown in Table \ref{tab:variance_matrix_estimates}. The ``classical'' estimate of the variance matrix is based on the negative inverse of the Hessian, while the robust variance matrix is calculated by ``sandwiching'' $\hat{O}$ between $\hat{H}^{-1}$. Finally, the BHHH estimate \citep{RePEc:nbr:nberch:10206} may be attractive for computational reasons, but may differ substantially from the classical covariance matrix. 

\begin{table}[t!]
  \centering
  \begin{tabular}{@{} l l @{}}
    \toprule
    \textbf{Estimate}           & \textbf{Formula} \\ \midrule
    Classical                   & $\displaystyle
                                   \hat{\Omega}= -\textstyle{\frac{1}{N}}\,\hat{A}^{-1}
                                   \;=\;
                                   -\,\hat{H}^{-1}$ \\[4pt]
    Berndt–Hall–Hall-Hausman (BHHH)  & $\displaystyle
                                   \hat{\Omega}_{B}= \textstyle{\frac{1}{N}}\,\hat{I}^{-1}
                                   \;=\;
                                   \hat{O}^{-1}$ \\[4pt]
    Robust sandwich             & $\displaystyle
                                   \hat{\Omega}_{R}= \textstyle{\frac{1}{N}}\,
                                   \hat{A}^{-1}\,\hat{I}\,\hat{A}^{-1}
                                   \;=\;
                                   \hat{H}^{-1}\,\hat{O}\,\hat{H}^{-1}=\hat{\Omega}\;\hat{\Omega}_{B}^{-1}\;\hat{\Omega}$ \\
    \bottomrule
  \end{tabular}
  \caption[Variance–covariance matrix estimators]{Standard variance–covariance matrix estimators.}
  \label{tab:variance_matrix_estimates}
\end{table}

\textbf{Hereafter, when not specifically stating which estimator is used for $\Omega$, the notation $\hat{\Omega}$ relates to the classical covariance matrix.}

If the model is correctly specified, and the optimum is well defined, the sampling variance of the MLE, i.e., $\hat{\Omega}$, will correspond to the Cram\'{e}r-Rao lower bound, i.e. the estimates are minimum variance. From $\hat{\Omega}$, we can obtain estimated standard errors as the square roots of its diagonal elements, say $\hat{\sigma}_k=\sqrt{\hat{\Omega}[k,k]}$ for parameter $\beta_k$. 

\subsection{Robust covariance matrix and bootstrapping}

A number of subtleties arise in relation to the earlier discussions. First, the properties discussed above rely on the notion that the model itself is correctly specified, which is in most cases a dubious claim to make at best. For example, if an analyst uses a linear specification when the true specification is non-linear, then the true values for our model parameters could not be obtained due to endogeneity \citep{guevara2024endogeneity}. By using ML estimation, the analyst will obtain \emph{consistent} estimates of the parameters for the \emph{specified model}, but these estimates, and their standard errors, will in general be affected by endogeneity if the specified model is not the correct model. In other words, the wrongly specified model has wrong estimates and wrong statistical significance properties. The standard error for $\hat{\beta}$ simply reflects the uncertainty in our estimated values for the estimated model, and will in general be wrongly estimated also. They are not related to any inconsistency caused by choosing the \emph{wrong} model. This already means that a wrongly specified model can lead to wrong inferences about statistical significance. One specific example of misspecification is the inclusion of irrelevant, or exclusion of relevant (i.e.\ missing), variables. This kind of misspecification not only influences the parameter estimates and potentially causes inconsistency, but also affects the size of the standard errors of all the parameters in the model due to its impact on the information matrix. Second, the properties relate to a distribution of the MLE around the \emph{true} parameter value for the \emph{population}, not for the \emph{sample}. The assumption that the model estimation yields the \emph{true} value for the parameter for the \emph{sample}, conditional on the model specification, is implicit. 

Under mild specification mistakes that do not compromise consistency, such as in some cases neglected heterogeneity or correlation between observations, two key approaches exist that make fewer assumptions than the classical covariance matrix calculations and thus provide a less conservative estimate of the sampling error.

The first approach is to use \emph{robust} standard errors, which, as discussed above, are obtained using the so called \emph{sandwich} estimator\footnote{The origin of the term \emph{sandwich estimator} is that $\Omega$ forms the bread, while the BHHH matrix is the filling.}, with:

\begin{equation}    
\hat{\Omega}_{R}=\hat{\Omega}\;\hat{\Omega}_{B}^{-1}\;\hat{\Omega},
\end{equation}
where we refer the reader back to Table \ref{tab:errors} for a definition of the individual components. If the model is correctly specified, asymptotically $\hat{\Omega}_B=\hat{\Omega}$, in which case the robust and classical covariance matrices are equal. In general, larger differences between classical and robust standard errors for specific parameters indicate a higher degree of misspecification for these parameters. Our personal experience is that, notwithstanding the general expectation that robust standard errors are larger than classical standard errors, the opposite can happen in some cases. Specifically, when robust standard errors are smaller than classical standard errors, this may indicate substantive misspecification or identification issues. Unfortunately, we do not have analytical proofs for these observations, but we refer the interested reader to \citet{King_Roberts_2015} to learn more about the extent to which robust standard errors identify methodological problems. Of course, it should be noted that robust standard errors do not fix these specification issues themselves, and this remains the task of the modeller. As the \emph{true} model form is never known, it would thus be wise to continue using robust standard errors even after making improvements to the model on the basis of the discrepancy between classical and robust standard errors.

The second approach is to use non-parametric methods, like \emph{bootstrapping}. This works by repeated sampling, and is the approach with least assumptions, but is computationally most demanding, meaning it is rarely used, especially for complex models. Bootstrapping recognises that the source of uncertainty is sampling error, and approximates this by simulating the sampling process with the data at hand. In particular, we would draw $S$ versions of the data with replacement (i.e.\ some people will be included multiple times) from the original sample, producing samples of the same size as the original data. The specified model would then be estimated on each of the samples, yielding $S$ sets of estimates of $\beta$, say $\hat{\beta}^{\left(s\right)}=\left<\hat{\beta}_{1}^{\left(s\right)},\hdots,\hat{\beta}_{K}^{\left(s\right)}\right>,\,s=1,\hdots,S$. If $S$ is sufficiently large, we obtain the result that:
\begin{equation}
\frac{\sum_{s=1}^S\hat{\beta}_{k}^{\left(s\right)}}{S}\rightarrow\hat{\beta}_k,
\end{equation}
where $\hat{\beta}_k$ is our original MLE on the collected data, see Equation (\ref{eq:mle}).

We then finally have that:
\begin{equation}\label{eq:boot_se}
\hat{\Omega}_{bootstrap}=var\left(\hat{\beta}^{\left(1\right)},\hdots,\hat{\beta}^{\left(S\right)}\right)    
\end{equation}

\subsection{Error measures for parameter transformations}\label{sec:delta}

While parameters from choice models provide initial insights into the studied behaviour, the key interest lies in transformations of such parameters. This includes most prominently marginal rates of substitution (MRS), such as willingness to pay (WTP). For simple linear specifications, these are given by ratios of parameters, while the computation becomes more difficult in the presence of non-linearity. Another reason for the need for transformation of parameters relates to the issue of effect sizes. Parameter values themselves do not tell us about the size of an effect in terms of impact on behavior, and neither do MRS measures, or indeed relative attribute importance (RAI). Instead, analysts need to look at demand predictions or elasticities \citep[cf.][]{repec:elg:eechap:20188_26}, or marginal effects \citep[see e.g.][]{HESS2022114800}. Independently of which specific measure an analyst is thus using for further interpretation, it should be clear that these measures themselves are affected by uncertainty, given that they are derived on the basis of parameters with uncertainty.

As discussed in detail by \citet{Daly2012a}, error measures can be computed for any functions of different model parameters using the Delta method, with the same underlying MLE properties applying to those error measures. \citet{DALY2023103828} show that the Delta method can also be applied using the sandwich matrix to obtain robust estimates of errors in functions of estimated parameters. Our discussions in terms of confidence intervals in Section \ref{sec:confidence_intervals} and hypothesis testing in Section \ref{sec:significance_tests} thus apply in the same way to functions of parameters as to the estimates of the parameters themselves. 

Additional complexities arise in the case of functions of randomly distributed coefficients. Let us assume that $\beta_1$ is allowed to vary randomly across individuals in a model. In the case of a continuous mixture model such as Mixed Logit, an analyst would obtain estimates of parameters that characterise the shape of the distribution of $\beta_1$, such as a mean and standard deviation. These parameters have uncertainty attached to them, thus also implying uncertainty in the extent of heterogeneity and shape of the distribution. We can then report uncertainty in these moments of the distribution, but this is not to confused with heterogeneity in $\beta_1$. 

If we now turn to metrics computed on the basis of distributed coefficients, such as for example a MRS with $\frac{\beta_1}{\beta_2}$, when the numerator and denominator are distributed, the analyst needs to be particularly mindful of the difference between uncertainty, which relates to estimated parameters, and heterogeneity, which relates to the variation of sensitivities in a population. Model estimation provides parameters that characterise the shape of the latter, and these parameters have uncertainty attached to them, thus also implying uncertainty in the extent of heterogeneity and shape of the distribution.

In the case of both continuous mixture models such as Mixed Logit and discrete mixtures such as Latent Class, an analyst can compute moments such as mean and variance for the distribution of measures such as $\frac{\beta_1}{\beta_2}$. If these moments have a closed form expression, as they do in Latent Class and for some choices of distribution of $\beta_1$ and $\beta_2$ (such as Lognormal) for continuous mixtures, then the Delta Method can be applied to calculate measures of uncertainty for such moments. However, for distributions where the ratio $\frac{\beta_1}{\beta_2}$ does not have defined moments, or where an analyst wants to produce a measure of uncertainty around the entire continuous distribution of $\frac{\beta_1}{\beta_2}$, rather than just the moments thereof, more complex approaches are needed that involve a level of numerical simulation, as discussed in \citet{repec:eee:transb:v:58:y:2013:i:c:p:199-214,SCACCIA202354}. 

\subsection{Repeated choice data}

Our discussion so far has focussed on a cross-sectional context, with one choice situation per person. A large number, and possibly a majority, of applications of choice models however make use of repeated choice data, i.e. with multiple observations per individual, with likely correlation between them. 

While a large share of studies explicitly model such correlations, for example by integrating over random distributions at the person (rather than observation) level \citep[cf.][]{Revelt1998}, this is not the case in work that relies on closed form models such as multinomial or nested logit. In such applications, post-estimation correction approaches are required. Indeed, the standard errors are likely to be underestimated, given that $T$ observations each from $N$ people is not the same as $1$ observation each from $NT$ people

Early correction approaches involved the use of Jackknife or Bootstrapping \citep[cf][]{760}, but a more straightforward approach is to rely on a panel implementation of the sandwich or bootstrap covariance matrix. 

For the sandwich approach, the typical implementation is for the summation in the BHHH matrix to be across individual people rather than choice observations, i.e.\ as in Equation (\ref{eq:BHHH}). This ensures that the calculation accounts for correlation across choices for the same person that the modeller may have omitted from the model specification. In such a case with repeated choice data, one can expect robust standard errors to be larger than classical standard errors because of the inherent correlation in choices across individuals not recognised by the classical standard errors.

A similar correction is possible by performing bootstrapping at the level of individual decision-makers as opposed to individual choice observations. Analysts should be mindful that the use of either robust standard errors or bootstrapping does not imply that their model has captured the repeated choice nature of the data, but simply that a correction of the standard errors has been applied. It is of course always preferable to explicitly account for such correlations during model estimation, if empirically and computationally feasible.

\subsection{Identification and over-specification issues}

Thus far, we have worked on the assumption that an analyst obtains an estimate of the covariance matrix after model estimation. There are however situations where that is not the case. If information captured in the data is insufficient to uniquely determine the model parameters, then the choice model is unidentified and the Hessian matrix $H$, and consequently the Fisher information matrix $I$, is singular. This means that the matrix has a determinant of zero and it cannot be inverted. As a result, the covariance matrix $\Omega$, which requires inverting the Hessian and/or Fisher information matrix, does not exist. There can be several reasons for a singular matrix, including having too many free parameters in the model (e.g., trying to estimate an alternative-specific constant for each alternative), or perfect collinearity between some attributes. An unidentified model can be made identified by normalising some of the parameters to reduce the number of free parameters, or by changing the structure of the model, for example by applying nonlinear transformations to independent variables or by including certain independent variables as an interaction instead of a main effect.
 
In addition to these theoretical identification issues, there is also scope for empirical identification issues, for example when including an interaction for a segment of people that always or never choose a given alternative. The associated parameter would in these cases tend to plus or minus infinity, respectively, and would not be empirically identified. Similarly, even in the absence of perfect collinearity, highly correlated independent variables can make the Hessian or information matrix almost singular, resulting in an ill-conditioned matrix with an unstable inverse and hence resulting in an unreliable covariance matrix.  
 
Analysts should also be aware of two additional caveats. First, there can be situations where even if a model is theoretically not identified, the issue can be hidden by the estimation process that is used. This is highlighted by \citet{957} in the context of simulation based estimation of mixture models, where the use of too low a number of draws can hide identification issues. Conversely, there can also be cases where, even though a model is theoretically and empirically identified, the complexity of the model leads to numerical issues in the computation of the Hessian. The upshot from these two points is that an analyst needs to take care in two directions. First, a user cannot be sure that the model is identified just because a covariance matrix is returned by the estimation software, and due diligence is thus needed in model specification. Second, when no covariance matrix is returned, the analyst should first check for obvious identification issues before considering whether numerical issues might also be a cause of problems.

Finally, analysts should also be mindful of a related issue, namely the risk of convergence to local optima that are (potentially very) inferior to the global optimum. This issue is more prevalent than most analysts might suspect \citep[cf.][]{hess2025holeprofilelikelihoodapproach}, and thus of course also adds some caveats to the faith analysts should place in the standard errors produced by their estimation software. Specifically, inference may reflect curvature at a local optimum rather than global likelihood geometry. Given the resulting possibility of substantially different estimates and standard errors, this could mean not only different effect sizes but a likelihood that with one local optimum, we find an effect, while with another, we do not, leading to different conclusions. A detailed discussion of this topic is beyond the scope of the present paper, but analysts are encouraged to invest time in investigating the use of different starting values, robust estimation approaches and targeted solutions \citep[see e.g.][]{Bierlaire2010}.

\section{Confidence intervals}\label{sec:confidence_intervals}

Confidence intervals (CIs) are widely used in reporting results and to indicate the uncertainty in the estimated values. However, CIs are also often poorly understood. Formally, a CI is a range of values where we can have a specified level of confidence that this contains the true value of an unknown population parameter. The subtlety here is that the population value is fixed, it is not a random variable. It is instead the CI which is varying between samples. For example, the interpretation of a $95\%$ CI for $\beta_k$ is that, if we were to repeat the data collection and analysis $100$ times and compute $95\%$ CIs for each of these, then approximately $95$ of them would contain the true parameter value $\beta_k^*$.

\begin{figure}[t!]
\caption{Graphical representation of the calculation of asymptotic confidence intervals}
	\begin{center}
	\includegraphics[width=14cm]{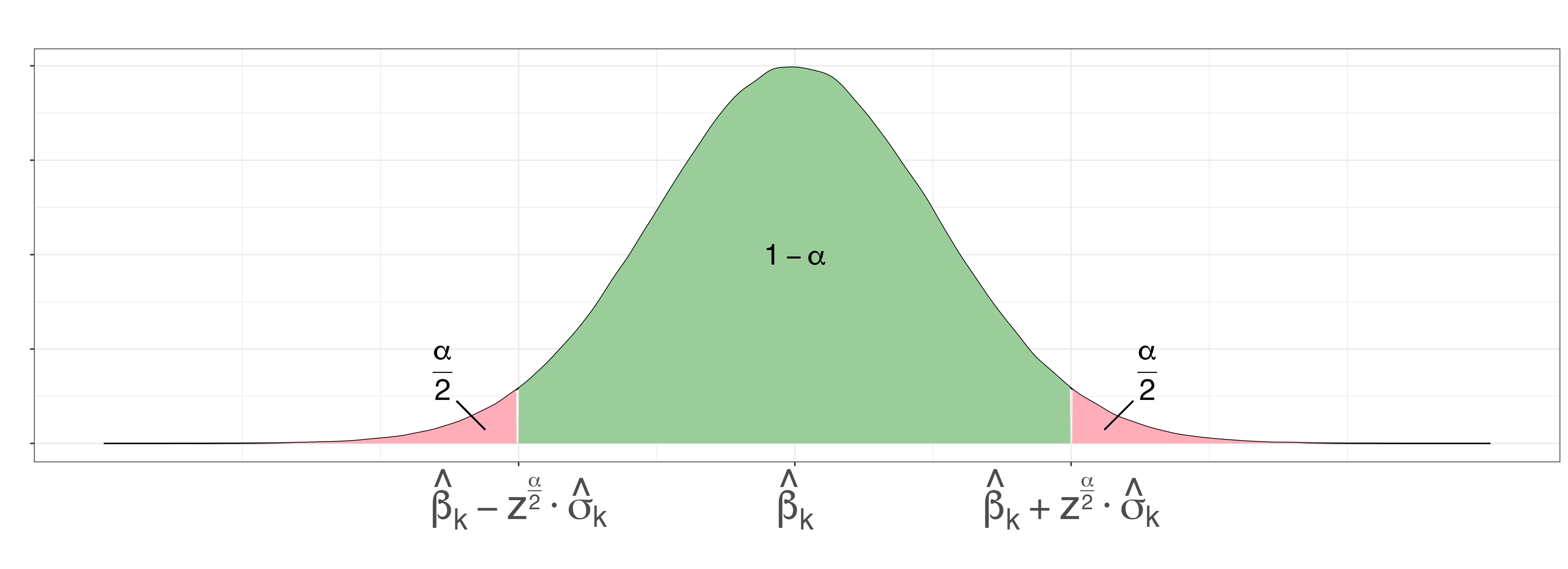}	
	\end{center}
	\label{fig:confidence}
\end{figure}

The standard way of computing CIs is to use asymptotic MLE properties, giving us asymptotic CIs. Formally, a $C$\% confidence interval for $\beta_k$ would be obtained as:
\begin{equation}\label{eq:CI}
\hat{\beta}_k\pm z^{\frac{\alpha}{2}}\cdot\hat{\sigma}_{k},
\end{equation}
where $\alpha=1-\frac{C}{100}$, such that $z^{\frac{\alpha}{2}}$ is the upper $\frac{\alpha}{2}$ critical value for a $\mathcal{N}(0,1)$ distribution, e.g., using $\hat{\beta}_k\pm 1.96\cdot\hat{\sigma}_{k}$ for a $95\%$ confidence interval. This process is explained in Figure \ref{fig:confidence}, which highlights the property of asymptotic normality. The link to the asymptotic normality assumption is clear; with a standard Normal distribution, $\frac{\alpha}{2}$ of the distribution lies below $-z^{\frac{\alpha}{2}}$, while $\frac{\alpha}{2}$ of the distribution lies above $+z^{\frac{\alpha}{2}}$. The addition of $\hat{\beta}_k$ and multiplication of $z^{\frac{\alpha}{2}}$ by $\hat{\sigma}_{k}$ in Equation (\ref{eq:CI}) simply moves us from a standard Normal distribution to the asymptotic Normal distribution for the MLE of $\beta_k$.

The calculation of asymptotic CIs in this manner is possible for individual parameters, or the computed value of functions of multiple parameters, using standard errors obtained via the Delta method \citep[cf.][]{Daly2012a}. If analysts wish to consider confidence intervals around the actual distribution of a random coefficient (or ratio of coefficients), then they further need to consider the work of \citet{repec:eee:transb:v:58:y:2013:i:c:p:199-214,SCACCIA202354}.

In computing confidence intervals, analysts can use classical standard errors, robust standard errors or bootstrap standard errors in Equation (\ref{eq:CI}). The key question that arises at this stage, however, is the impact of the assumptions underlying the computation of asymptotic confidence intervals.

To be specific, if the estimated parameters were distributed exactly normal, this would imply that the log-likelihood function was quadratic in the parameters. A shift away from the optimum $\hat{\beta}_k$ for parameter $\beta_k$ by a value $\Delta_{\beta_k}$ would reduce the log-likelihood by $\Delta LL=\frac{{\Delta_{\beta_k}}^2}{2{\hat{\sigma}_{k}}^2}$. However, the `asymptotic' qualifier means that the property applies exactly only as the sample size tends towards infinity. For finite data sets, it is an approximation that applies only in the neighbourhood of the optimum, i.e. for small $\Delta_{\beta_k}$. Indeed, as is shown by \citet{406}, for example, the log-likelihood function is not exactly quadratic, so that reductions in likelihood do not follow this formula exactly at all. What happens as we move further away from the optimum is not defined by maximum likelihood theory. The parameter estimates are not distributed exactly normal, and any theoretical insights/claims based on an assumption of normality are misguided. The problem is that what defines \emph{``close''} is unclear and depends on the model. There is thus no guarantee that the shape of a $C$\% confidence interval (especially for large $C$) is normal, that the bounds are at or close to $\hat{\beta}_k\pm z^{\frac{\alpha}{2}}\cdot\hat{\sigma}_{k}$ or that the bounds are symmetric around the MLE, $\hat{\beta}$.

If an analyst wishes to compute confidence intervals without relying on these assumptions, then an alternative approach is to rely on bootstrapping, whether looking at individual parameters, or functions of parameters (as in \citealt{999}). We earlier explained how bootstrapping can be used to produce bootstrap standard errors (cf.\ Equation (\ref{eq:boot_se})). Rather than using bootstrap standard errors in an asymptotic confidence interval, an analyst can produce an empirical confidence interval. Remember that bootstrapping is based on producing $S$ different estimates of $\beta_k$ through resampling. The $C\%$ bootstrap confidence interval can be computed using the empirical distribution of these estimates. Denote the bootstrap estimates as $
\{\hat{\beta}_k^{(1)}, \hat{\beta}_k^{(2)}, \ldots, \hat{\beta}_k^{(S)}\}$. The lower $(L_{\beta_k})$ and upper $(U_{\beta_k})$ limits of the $C\%$ bootstrap CI are given by the quantiles of the bootstrap estimates. The resulting confidence interval is generally not symmetric around $\hat{\beta}_k$, reflecting the empirical distribution of the bootstrap estimates. Just as for bootstrap standard errors, the computation of bootstrap confidence intervals relies on an analyst decision on the number of bootstrap samples. More robust results will be obtained with a higher value for $S$, albeit at a higher computational cost. This aside, if an analyst wishes to use empirical bootstrap confidence intervals, it is wise to choose a value of $S$ for which $\frac{\alpha}{2}\cdot S$ is an integer.

Since the empirical distribution of parameter estimates obtained by means of bootstrapping is not necessarily symmetric, the use of quantile functions can potentially result in a slight misrepresentation of the confidence intervals. That is, quantile functions typically indicate the lowest and highest 2.5\% of the empirical distribution to denote the 95\% confidence interval. This however ignores that there may be a narrower support (or interval) which contains 95\% of the density of interest. For example, a positive lognormal density interval has a large density close to zero and a very long tail. We can thus find a support area closer to zero which contains 95\% of the density and thereby provide a more accurate representation of the confidence interval. This is often referred to as a Highest Posterior Density (HPD) Interval \citep[see e.g.][]{gelman2013bayesian}\footnote{As an example, assuming a lognormal density with a mean of zero and standard deviation of one for the underlying normal density, provides a 95\% confidence interval between 0.14 and 7.1 using the quantile function and a HPD interval between 0.03 and 5.18, where the latter is clearly much narrower than the former. With larger sample sizes, the standard errors will decrease, and these alternative definitions of confidence intervals are expected to converge due to the assumption of asymptotic normality.}. This functionality is commonly implemented in packages such as `Coda' \citep{coda}.     

\section{Hypothesis tests}\label{sec:significance_tests}

Choice modellers use hypothesis testing for two key purposes, namely tests on individual parameters (or functions of parameters), and tests to compare different model specifications against each other. As we will see in this section, there is a strong relationship between these two types of tests, which is often misunderstood.

\subsection{Formulation of hypotheses and possible test outcomes}\label{sec:hypotheses}

In statistical tests, an analyst forms two hypotheses, a null hypothesis, $\mathbf{H_0}$, and an alternative hypothesis $\mathbf{H_1}$. The test process then involves determining whether there is sufficient evidence to reject $\mathbf{H_0}$ in favour of $\mathbf{H_1}$.

%In Section \ref{sec:hypotheses}, we discussed the process of formulating a null and alternate hypothesis, while in Section \ref{sec:tests}, we described the three key tests used to decide whether the null hypothesis $\mathbf{H_0}$ can be rejected given the evidence found for $\mathbf{H_1}$. 

As shown in Table \ref{tab:errors}, there are four possible cases that combine the \emph{truth} in the data with the \emph{decision} by the analyst. The analyst does not know whether $\mathbf{H_0}$ is true or false, and thus needs to determine what risk of making a wrong decision is acceptable. The core emphasis in statistical tests conducted in choice modelling relates to how confident we can be in rejecting $\mathbf{H_0}$, and in particular what the risk of a type I error is, i.e. rejecting the null hypothesis even though it is true. When conducting a statistical test, we obtain a $p$ value for the given test statistic, which is the probability of observing a value as high as the test statistic when $\mathbf{H_0}$ is true. In simpler terms, the $p$-value is the probability that our finding in support of rejecting $\mathbf{H_0}$ was obtained by chance. A smaller $p$-value suggests stronger evidence to reject $\mathbf{H_0}$. An analyst needs to set the maximum acceptable probability of making a Type I error, where this is defined as $\alpha_{\text{type I}}$, also known as the significance level of the test\footnote{We add the subscript in $\alpha_{\text{type I}}$ to avoid any confusion with model parameters defined as $\alpha$.}. The $p$-value obtained via a statistical test is compared to $\alpha_{type I}$, and $\mathbf{H_0}$ is rejected if $p\leq\alpha_{type I}$. Setting a lower $\alpha_{type I}$ thus reduces the risk of a type I error. 

\begin{table}[t!]
\centering
\renewcommand{\arraystretch}{1.5} % Adjust row height for better readability
\begin{tabular}{@{}lcc@{}}
\toprule
& \textbf{Accept $\mathbf{H_0}$} & \textbf{Reject $\mathbf{H_0}$} \\ \midrule
$\mathbf{H_0}$\textbf{ is true} & 
\begin{tabular}{c} 
True negative \\ 
(Prob = \( 1 - \alpha_{\text{type I}} \)) 
\end{tabular} & 
\begin{tabular}{c} 
Type I error / False positive \\ 
(Prob = \( \alpha_{\text{type I}} \)) 
\end{tabular} \\ \midrule
$\mathbf{H_0}$\textbf{ is false} & 
\begin{tabular}{c} 
Type II error / False negative \\ 
(Prob = \( \beta_{\text{type II}} \)) 
\end{tabular} & 
\begin{tabular}{c} 
True positive \\ 
(Prob = \( 1 - \beta_{\text{type II}} \)) 
\end{tabular} \\ \bottomrule
\end{tabular}
\caption{Error Types in Hypothesis Testing}\label{tab:errors}
\end{table}

A topic that has received far less attention in the choice modelling literature, with the exception of work on experimental design \citep[cf.][]{pub:81584} and tests for endogeneity \citep{guevara2018overidentification}, is that of the power of statistical tests, which relates to the risk of type II errors. An analyst does not know whether $\mathbf{H_0}$ is true or not, but wants to determine the likelihood that the test will provide the correct recommendation when $\mathbf{H_0}$ is false and $\mathbf{H_1}$ is true. The issue is that $\mathbf{H_1}$ is compound, meaning it represents an interval of infinitely many possible values, such as $\mathbf{H_1}: \beta_k<0$. An analyst can however calculate the ``power'' of the test for a given ``effect size'', which is a specific value of the alternative hypothesis. A Type II error occurs when $\mathbf{H_0}$ is false, but the test fails to reject it, and this type of error has a probability of $\beta_{\text{type II}}$. The power of a test is given by $1-\beta_{\text{type II}}$ and represents the probability of rejecting $\mathbf{H_0}$ when a specific alternative hypothesis is true\footnote{We add the subscript in $\beta_{\text{type II}}$ to avoid any confusion with model parameters defined as $\beta$.}. As the effect size increases, the power of the test also increases because it becomes easier to correctly reject $\mathbf{H_0}$. The power of a test can also be used to evaluate and compare the effectiveness of different tests. A test with higher power for a specific alternative hypothesis is preferred, as it is more likely to correctly reject $\mathbf{H_0}$. 

As described above, a statistical test is used to determine whether the data support $\mathbf{H_1}$ over $\mathbf{H_0}$, i.e. allow us to reject $\mathbf{H_0}$. There are some important nuances that are often misunderstood. First, rejecting $\mathbf{H_0}$ does not mean accepting $\mathbf{H_1}$. It simply means that $\mathbf{H_0}$ is unlikely to be true, given the evidence in support of $\mathbf{H_1}$. Second, a failure to reject $\mathbf{H_0}$ does not mean that $\mathbf{H_0}$ is true, but simply means that we do not have strong enough evidence to reject $\mathbf{H_0}$. Formally, this means that analysts should not make statements that they have evidence that an effect exists, or even that they are testing whether an effect exists. Indeed, the correct statement would be that they can reject the hypothesis that an effect does not exist. Of course, from a practical perspective, rejecting $H_0$, i.e. observing an effect so large that it would be unlikely under the null hypothesis, can reasonably be interpreted as an indication that such an effect exists, even though this does not allow us to ascribe an unconditional probability to the existence of that effect. Another related point concerns was \citet{Ziliak} call the fallacy of the transposed conditional. A statistical test considers the probability of observing the data given that $\mathbf{H_0}$ is true, expressed as the $p$-value. This is not the same as the probability that $\mathbf{H_0}$ is true given the data.

The specification of hypotheses differs depending on whether an analyst conducts tests for individual parameters or tests comparing different models, as we will discuss in the following sections.

\subsection{Hypotheses for tests on individual parameters}\label{sec:H_params}

When using statistical tests for individual parameters, our key interest is in whether a given parameter differs from a specific value. We first formulate a null hypothesis for parameter $\beta_k$ as:
\begin{description}
    \item[$\mathbf{H_0}$:] $\beta_k = \beta_{k,H_0}$
\end{description} 
Notwithstanding the discussions in the introduction about the widespread question of how valid it is to consider the possibility of a parameter having a value of zero, the most common value for $\beta_{k,H_0}$ in this case is indeed zero\footnote{For some parameters, such as multipliers, scale parameters, or nesting parameters, the useful test is one where $\mathbf{H_0}: \beta_k = 1$.}. There are of course parameters where it is entirely reasonable to consider the possibility that the estimate is zero. For many others, such as key behavioural attributes like time and cost, this is more doubtful. Even here though, reasons exist, for example the possibility that in the case of poor data or inappropriate design for a stated choice survey, the attribute has little or no possibility of affecting behaviour. In this case, an analyst will want to be able to test this, which means that the test is used to determine whether $\beta_k$ has an effect in the model. It is important to recognise that, formally, we do not establish whether an effect exists, but whether we can reject the null hypothesis that no effect exists. Specifically, we thus set: 
\begin{description}
    \item[$\mathbf{H_0}$:] $\beta_k = 0$
\end{description} 
Alongside the null hypothesis, there is a requirement for the definition of the alternate hypothesis, i.e. $\mathbf{H_1}$. With generally limited discussion in published work, the default assumption is to set this as:
\begin{description}
    \item[$\mathbf{H_1}$:] $\beta_k \neq 0$,
\end{description} 
which equates to using a two-sided test. For confidence intervals, it clearly makes sense to work with a two-sided confidence interval, looking at the distribution to either side of the estimated value. However, a different reasoning should apply in hypothesis testing. Indeed, for the majority of parameters used in choice models, there is a strong a priori sign assumption. For example, a cost coefficient is a priori assumed to be non-positive. A model that produces a positive cost coefficient would be rejected on the basis of counter-intuitive results. This in turn means that, when such a strong sign assumption exists, this should be reflected in the formulation of $\mathbf{H_1}$. For example, if we a priori assume that $\beta_k$ should be non-positive, then we would use:
\begin{description}
    \item[$\mathbf{H_0}$:] $\beta_k = 0$,
    \item[$\mathbf{H_1}$:] $\beta_k < 0$,
\end{description} 
which equates to using a one-sided test. 

Many papers report the outcomes of such tests without indicating if a one-sided or two-sided test was used, and the default assumptions in many software packages is two-sided\footnote{Apollo \citep{hess_palma_apollo} uses one-sided tests by default but gives the user the option to request two-sided tests instead.}. The inappropriate use of a two-sided test has two implications. First, it means that the definition of a type I error would cover two areas, namely that where a parameter is positive, or where it is very negative. Second, in terms of actual findings, the use of a two-sided test will produce $p$-values that are double those of a one-sided test, increasing the risk of type II errors.

A special case that warrants some attention is for a parameter that has a theoretical requirement to be contained between certain values, such as a structural parameter ($\lambda$) in a nested logit model that needs to be between $0$ and $1$. A common test reported in papers is with:

\begin{description}
    \item[$\mathbf{H_0}$:] $\lambda = 1$,
    \item[$\mathbf{H_1}$:] $\lambda \neq 1$,
\end{description} 

\noindent which would again be a two-sided test, which would mean that values that are far below the estimate, but potentially still above $0$ would also be unacceptable. In such a situation, it would be safer for an analyst to instead conduct two one-sided tests, namely

\begin{description}
    \item[$\mathbf{H_0}$:] $\lambda = 1$,
    \item[$\mathbf{H_1}$:] $\lambda < 1$,
\end{description} 

\noindent and

\begin{description}
    \item[$\mathbf{H_0}$:] $\lambda = 0$,
    \item[$\mathbf{H_1}$:] $\lambda > 0$.
\end{description} 

\noindent As an alternative to conducting two one-sided tests, an analyst could also directly work out the $p$-value as being the probability of the estimate falling outside the $\left[0,1\right]$ interval, i.e. what part of a $N\left(\hat{\lambda},\hat{\sigma}_{\lambda}\right)$ falls outside this interval, where $\hat{\lambda}$ and $\hat{\sigma}_{\lambda}$ are the MLE and standard error, respectively.

\subsection{Hypotheses for tests comparing models}\label{sec:model_comparison}

The second use of statistical tests relates to comparing two model specifications against each other. Our focus here is on formal statistical tests that require one model, say $M_r$, to be a restricted version of a more general model, say $M_g$.

The null hypothesis is then: 
\begin{description}
    \item[$\mathbf{H_0}$:] $LL_{M_g}=LL_{M_r}$.
\end{description} 
Given that the general model nests the restricted model, by definition, its log-likelihood will be less negative, and we thus formulate the alternate hypothesis as:
\begin{description}
    \item[$\mathbf{H_1}$:] $LL_{M_g}>LL_{M_r}$,
\end{description} 
meaning that a one-sided test will be used.

\subsection{A trinity of tests}\label{sec:tests}

We will now focus on the relationship between the tests used for constraints on individual parameters and tests for comparing model specifications. In maximum likelihood estimation, there is a \emph{trinity} of tests that are asymptotically equivalent to each other, i.e. lead to the same findings when the sample size becomes sufficiently large. 

To discuss the differences between these three tests, and their use in choice modelling, we focus on the case of a constraint on a single parameter - extension to multiple parameter constraints are relatively straightforward.

Let us again assume we have a parameter of interest $\beta_k$, for which we formulate 

\begin{description}
    \item[$\mathbf{H_0}$:] $\beta_k = 0$,
    \item[$\mathbf{H_1}$:] $\beta_k \neq 0$.
\end{description} 

\noindent We can estimate a constrained model (i.e. reflecting $H_0$), which gives us a log-likelihood $LL_{H_0}$ and a general model which freely estimates $\beta_k$, giving us $LL_{H_1}$ with an estimate $\hat{\beta}_k$ for the parameter in question. The $H_1$ model thus estimates one additional parameter compared to the $H_0$ model.

The three tests of interest are:

\begin{description}
    \item[Likelihood ratio (LR) test:] The LR test compares the fit of a general model and its constrained version. If $H_0$ is true, the test statistic converges asymptotically to a $\chi^2$ distribution with $d$ degrees of freedom, where $d$ is the number of additional parameters in the general model. Let $\hat{\beta}$ be the parameters obtained in the unconstrained estimation, while $\tilde{\beta}$ refers the constrained case, e.g. where we constrain $\beta_k=0$. We then have:
    \begin{equation}\label{eq:LR}
    LR=2\left(LL_{\hat{\beta}}-LL_{\tilde{\beta}}\right)\sim \chi^2_d,
    \end{equation}
    where, for our simple example, $d=1$. The name likelihood ratio test stems from the fact that the test is based on the logarithm of the ratio of likelihoods, which then of course gives the difference in log-likelihoods shown in Equation (\ref{eq:LR}). 
    \item[Wald test:] The Wald test involves estimating the general model, i.e. without parameter constraints. In the case of a single parameter, the Wald test, in a scalar version, is given by:
    \begin{equation}\label{eq:Wald}
    W=\frac{\left(\hat{\beta}_k-\beta_{k,H_0}\right)^2}{\hat{\sigma}_{k}^2}\sim \chi^2_1,
    \end{equation}
    where, with the above, $\beta_{k,H_0}=0$.
    
    Most choice modellers will be more familiar with the the square root of Equation (\ref{eq:Wald}), which gives us a (pseudo\footnote{Unlike in linear regression with normally distributed error terms, the $t$-ratio in our case does not actually follow a $t$-distribution. A $t$-ratio is also known as $t$-statistic or $t$-value. The term $t$-test should not be used when reporting $t$-ratios.}) $t$-ratio that asymptotically converges to a standard Normal distribution\footnote{The square of a standard Normal distribution is a $\chi^2$ distribution with one degree of freedom.} under $H_0$:
    \begin{equation}\label{eq:t_ratio}
    t=\frac{\hat{\beta}_k-\beta_{k,H_0}}{\hat{\sigma}_{k}}\sim \mathcal{N}\left(0,1\right)
    \end{equation}
    \item[Lagrange multiplier (LM) test\footnotemark:] The LM test involves estimating only the constrained model, giving a MLE vector $\tilde{\beta}$, i.e. in our case fixing parameter $\beta_k$ to a value of $\beta_{k,H_0}$. The LM test is then given by:
    \begin{equation}\label{eq:LM}
    LM=G_{\tilde{\beta}}^TI_{\tilde{\beta}}^{-1}G_{\tilde{\beta}}\sim \chi^2_d
    \end{equation}
    The test value uses the expectation of the score/gradient (i.e. first derivatives) and the Fisher information matrix (i.e. negative expectation of second derivatives of the log-likelihood function)  of the general model (i.e. without constraints) at the estimates of the constrained model. If $H_0$ is true, the test statistic converges asymptotically to a $\chi^2$ distribution with $d$ degrees of freedom, where $d$ is the number of additional parameters in the general model. 
\end{description}
\footnotetext{The LM test is also called score test as it is based on the slope of the log-likelihood function.}

\begin{figure}[t!]
    \centering
    \includegraphics[trim=0cm 1cm 0cm 1cm, clip, width=0.5\linewidth]{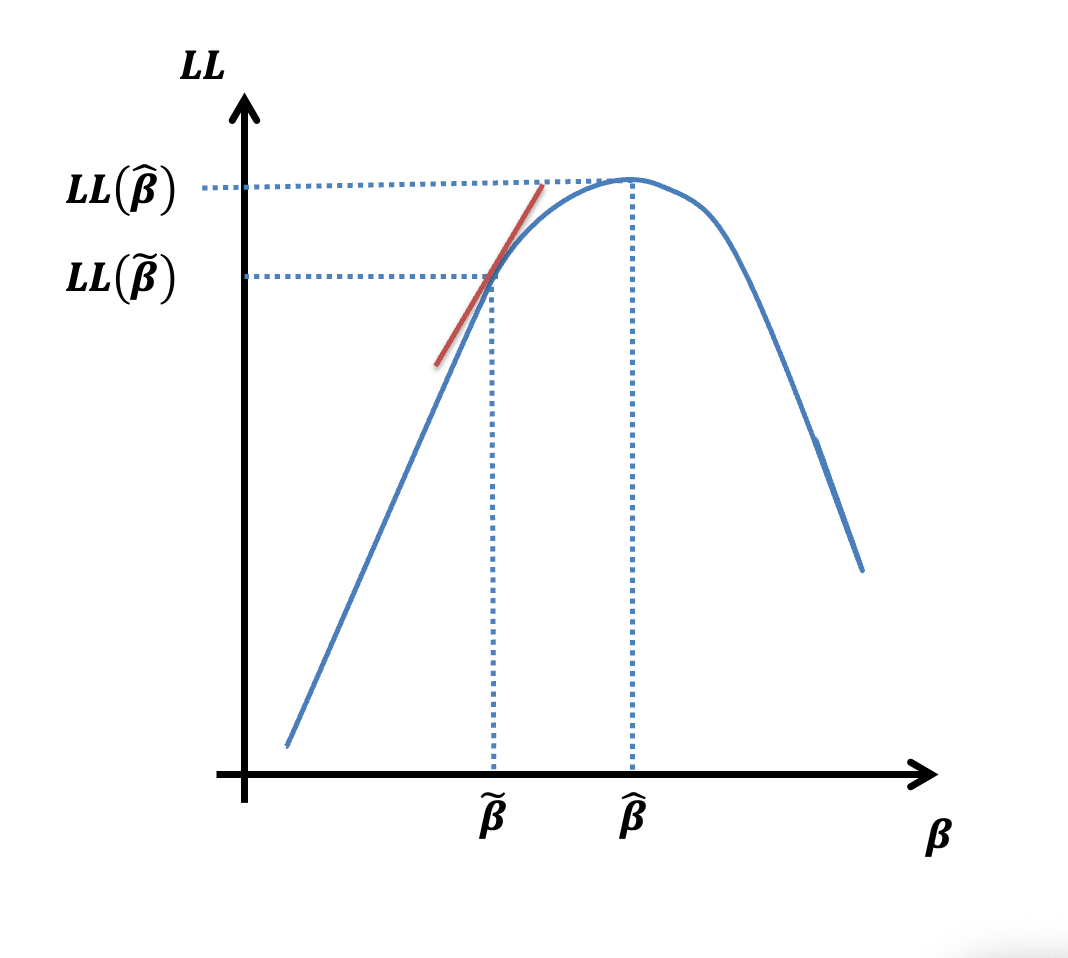}
    \caption{Relationship between estimates and log-likelihood values used for likelihood ratio, Wald, and Lagrange multiplier tests}
    \label{fig:trinity}
\end{figure}

\noindent To understand the relationship between the three tests described above, Figure \ref{fig:trinity} is helpful. Of the three tests, only the LR test uses information from both the general and constrained model by comparing $LL\left(\hat{\beta}\right)$ and $LL\left(\tilde{\beta}\right)$. The Wald test, and thus also the widely used $t$-ratios, uses information only from the general model, in terms of $\hat{\beta}$ and $\hat{\sigma}$. Finally, the LM test uses information only on the shape of the unconstrained log-likelihood function at the solution from the constrained model $\tilde{\beta}$, indicated by the red slope line. For the LR test, both models thus need to be estimated, for the Wald test, only the general model, and for the LM test, only the constrained model. The three tests are asymptotically equivalent but may diverge substantially with smaller samples. Overall, by using both log-likelihood values, the LR test does not make assumptions about the shape of the likelihood function, i.e. it does not rely on the property of asymptotic normality. If both the restricted and the unrestricted models are easy to evaluate, it should be preferred as it generally has more power \citep[see e.g.][]{Gudicha2016StatisticalPO}. This test is often preferred in DCM when assessing constraints that involve many coefficients together, since the likelihood of the restricted and the unrestricted model can be directly obtained from the model output, but in the context of behaviourally important individual parameters, it should potentially also be the preferred test when looking at individual parameters, especially if a simple t-ratio does not allow us to reject the hypothesis that the parameter is equal to zero. 

\subsection{Considerations specific to choice models}

Our discussion so far has looked at the topic from a broad perspective, and we now turn our attention to a few considerations that are specific to choice models. We again look separately at tests for individual measures, and at tests relating to model comparisons.

\subsubsection{Tests for derived measures}

The discussion in Section \ref{sec:H_params} and the details on the Wald test (and $t$-ratios in Section \ref{sec:tests}) focussed entirely on individual parameters. We now extend this to functions of parameters, and finally the incorporation of random heterogeneity.

The first extension is straightforward, given the properties of the Delta method, as discussed by \citet{Daly2012a}, and as highlighted in Section \ref{sec:delta}. An analyst can thus use the exact same formulation of hypotheses and tests discussed here for the case of measures computed on the basis of estimated model parameters, such as for example a MRS computation. 

The second extension, i.e. the presence of random heterogeneity, introduces additional complexity. Let us again focus on the case of a continuous mixture model, such that $\beta_1$ follows a distribution across individual people. The formulation of hypotheses in this case can take multiple forms. An analyst can conduct tests on the parameters representing the distribution, such as for example the mean and standard deviation. A test on the former would look at whether we can reject the hypothesis that the mean sensitivity is zero, while a test on the latter would allow us to test whether we can reject the hypothesis that there is no heterogeneity across individuals. This is however different from formulating a test on a distributed $\beta_1$, i.e. testing whether the population level distribution is different from zero. For this, an analyst needs to consider both the parameter uncertainty and the distribution of $\beta_1$ across people, for which we again refer the reader to \citet{repec:eee:transb:v:58:y:2013:i:c:p:199-214,SCACCIA202354}. %Of course, when a strict sign restriction is imposed, such as when using Lognormal distributions, the value of a statistical test against zero becomes moot.

\subsubsection{Model specification tests}

Choice modellers routinely estimate multiple different specifications on the same data and compare the resulting models against each other. This relates to situations where additional terms are added to a model, for example interactions between variables, treatments of non-linearity, heterogeneity, and more flexible assumptions about the error structure, notably in relation to correlation between the unobserved utility components for different alternatives.

To a large extent, such comparisons can make use of LR tests as they compare a restricted version of a model to a more general model. This includes for example models with additional interaction terms, the comparison of models with more flexible error structures to simpler models (e.g. Nested vs Multinomial Logit), or models with and without random heterogeneity (comparing Mixed Logit to Multinomial Logit). In each case, a simple parameter constraint would make the general model equivalent to the restricted model (generally setting concerned parameters to $0$, or in the case of nesting structures and multipliers to $1$)

At other times, a LR test is not possible, i.e. when one model is not a restricted version of the other model that can be explained by a simple parameter constraints. Some of these examples are well known, such as when comparing models with different types of non-linearity (e.g. linear \emph{vs} logarithmic), models with different treatments of heterogeneity (e.g. Mixed Logit \emph{vs} Latent Class) or models from different model families (e.g. Random Utility \emph{vs} Decision Field Theory). There are also examples that are less well recognised, and where analysts may mistakenly think they can apply a LR test. One example relates to comparing Latent Class models with different numbers of classes. A three class model can collapse to a two class model either by one class getting a zero class allocation probability, or by the parameters inside two classes collapsing to the same values. Another example relates to comparing a Cross-Nested Logit model to a Nested Logit model. Again, multiple restrictions potentially yield the simpler model, namely restrictions on the nesting parameters or the allocation parameters In both those cases, there are multiple ways in which a model can collapse into the other and hence the LR-test does not make sense, as it is not clear what the appropriate degrees of freedom would be. 

These situations relate to testing non-nested hypotheses. Informal comparisons can be made on the basis of goodness of fit statistics such as the adjusted $\bar{\rho}^2$ measure \citep[cf.][]{domencich1975urban, RePEc:eee:eejocm:v:21:y:2016:i:c:p:60-65}, or the Akaike Information Criterion \citep[AIC, cf.][]{akaike1974aic} and the Bayesian Information Criterion \citep[BIC, cf.][]{schwarz1978bic}. 

The adjusted $\bar{\rho}^2$ has special interest given that a formal test exists using it for the comparison of non-nested models. Let us define $\bar{\rho}^2=1-\frac{LL_{\hat{\beta}}-K}{LL_0}$, where $LL_{\hat{\beta}}$ and $LL_0$ are the log-likelihoods at convergence and for a model with equal probabilities for all available alternatives, respectively, while $K$ is the number of estimated parameters. \citet{ben1986akaike} determined a bound for the probability that a model may have a larger adjusted $\bar{\rho}^2$, while not being the one behind the data generation process, given by:
\begin{equation}
\texttt{Pr}(\bar{\rho}_1^2-\bar{\rho}_2^2\geq z)\leq \Phi\left[-\left(2\cdot z\cdot LL\left(\beta_{ES}\right)+df_1-df_2\right)^{\frac{1}{2}}\right],
\end{equation}
where
$z$ is observed difference in adjusted $\rho^2$, $\Phi$ is cumulative standard normal distribution, $LL\left(\beta_{ES}\right)$ is log-likelihood at equal shares, and $df_1$ and $df_2$ are the number of parameters for models 1 and 2.

This repertoire of LR tests, goodness of fit statistics, and the Ben-Akiva \& Swait test cover the majority of cases of model comparisons in choice modelling. The well known Swait \& Louviere test \citep{swait1993scale} for example uses a sequence of separate LR tests for comparing a generic pooled model to a model with scale differences and separate models. Similarly, comparing models that make different error structure assumptions to test for violations of the IIA property\footnote{Note this is different from comparing parameter estimates of models estimated with a full set of alternatives as opposed to a subset, as in the \citet{hausman1984specification} work, albeit that there also, statistical tests could then be used to compare estimates, at least in the form of MRS.} relies on LR tests \citep[cf.][]{brownstone1999forecasting}.

\subsection{Making decisions based on tests}\label{sec:test_outcomes}

As discussed earlier, statistical tests are used to decide whether there is enough evidence in support of $\mathbf{H_1}$ to reject $\mathbf{H_1}$. In this process, an analyst needs to balance the risks of type I and type II errors. A crucial step in the statistical test process is the decision the analyst makes for $\alpha_{\text{type I}}$, i.e. the selection of the significance level. It is very customary to set $\alpha_{\text{type I}}=0.05$, and many analysts use that level mechanically in their decisions of which parameters to retain in their models when looking at $t$-ratios, or to decide whether or not to reject a restricted model. Such a mechanical malpractice should be avoided. The selection of $\alpha_{\text{type I}}$ should depend on a mix of the plausibility of the null hypothesis and the cost of making a wrong decision. In classical hypothesis testing, the plausibility of the null hypothesis is very high, justifying small $\alpha_{\text{type I}}$, but this is not true for specification tests, such as the ones we face in model building.

While choosing a small value for $\alpha_{\text{type I}}$ may seem scientifically sound, as it reduces the risk of a Type I error, it also implies a higher probability of a Type II error. Imagine a decision about whether or not to remove a parameter from a model with a $p$ value higher than $\alpha_{\text{type I}}$. The consequence of a Type I error (erroneously including an attribute) can be a reduction in efficiency. Conversely, a Type II error (erroneously excluding an attribute) could cause endogeneity, invalidating the whole model. Especially in the case of variables where there is clear a priori knowledge that they should affect decisions, such as cost, excluding an attribute with a higher-than-usual $p$ value for the associated coefficient can be a serious mistake. Continuing with the cost example, it is very plausible that cost influences decisions and that excluding it from the model has a major impact on both estimation and forecasts made with the model; therefore, it is clear that cost should not be excluded from the model. Other cases are not so clear, which reinforces the importance of understanding what is being modelled, i.e., the irreplaceable knowledge that the researcher has of the phenomenon.

The finding of a large $p$-value for a variable could simply be the result of an insufficient sample size, low variance in the data or an incorrect utility specification (e.g. assuming linearity when the data generation process is different from this). An analyst would then be wise to address the situation by either collecting more data or seeking to improve the model, and if all else fails, retain the parameter but note its weak statistical properties. 

Excluding a variable is essentially assuming that the value zero is more reasonable than the value that was estimated, and that the cost of excluding it is not large enough, which is sometimes not true. It should be stressed that a parameter can be behaviourally meaningful and have a non-trivial impact on behaviour in a model even if it doesn't pass the $95\%$ threshold. Similarly, the inclusion of a parameter may be justified on the basis of it being an important policy-test input, providing the value is plausible \citep[see discussions in e.g.][chapter 8]{799}. Finally, when looking at niche applications with a limited pool of potential decision makers to include in the sample, the power of tests is further reduced, and less stringent significance levels may well deserve consideration and the a-priori expectations for the influence of a variable on choice become even more relevant. The decision about which parameters to retain when $p>0.05$, or equivalently which $\alpha_{\text{type I}}$ to use, is a multi-objective decision and highly context-dependent; analysts are advised to keep the above points in mind. But, as a general rule, in model building, i.e., specification testing, one may consider setting a universal upper bound for $\alpha_{\text{type I}}$, and in any case, it is always better to retain the parameter and report the $p$-value to make the results and limitations of the modelling transparent.

The tests carried out in choice modelling fall under the umbrella of specification tests, rather than ``classical tests'' where $\mathbf{H_0}$ is usually well-founded, and the cost of making a Type I is especially high. A common example is the criminal justice system, where the null hypothesis assumes innocence. Here, the cost of wrongly rejecting the null (convicting an innocent person) is very high. In the context of specification tests on the other hand, we are often uncertain about $\mathbf{H_0}$ or even suspect it should be rejected, such as when it relates to testing whether key parameters are equal to zero. In such cases, the cost of a Type II error (wrongly accepting the null) can be substantial, leading to model misspecification. Consequently, it is better to choose a larger $\alpha_{\text{type I}}$, increasing the power of the test and the likelihood of rejecting the null hypothesis when the alternative hypothesis is true.

Of course, the opposite argument can be made, i.e. making a case that $95\%$ confidence is actually a low level (meaning that a 5\% risk of a type I error is high). With the large datasets used in many studies on revealed preference (RP) data, or with good experimental designs for stated choice (SC) data, $t$-ratios in excess of $10$ or $20$ are often the rule rather than the exception, and $t$-ratios of $1.96$ (for a typical two-sided test) are thus low. Let us also consider real-world implications, and imagine a situation where a parameter estimate obtains a t-ratio against zero of $t_{0,\hat{\beta}_k}=4$, well into the territory of rejecting $\mathbf{H_0}: \beta_k=0$ at the $99\%$ confidence level. However, talking about it from a confidence interval perspective, it would imply that a 95\% confidence interval runs roughly from 50\% of the estimated value to 150\% of the estimate, which is not much use if we consider using the information to justify a multi-billion pound investment.

The point about $\alpha_{\text{type I}}$ being the significance of a test also highlights a common bad practice. Analysts routinely describe the findings of statistical tests as showing that a parameter is ``significantly different from $0$''. It should first be noted that hypothesis testing is only concerned with whether the null hypothesis can be rejected, not whether the alternative hypothesis is true. In other words, it establishes whether our finding could be due to chance and that the true value is in fact zero (if $\mathbf{H_0}$: $\beta_k = 0$). What is commonly described as a test of significance is thus a test to see whether the null hypothesis of a parameter being equal to zero can be rejected. The statistical significance of the test relates to the probability of $H_0$ being rejected when it is in fact true, i.e. a type I error. It is thus incorrect to talk about a parameter being $95\%$ significant. The significance level in this case is $5\%$, and we can reject $H_0$ with a $95\%$ level of confidence, not significance. The relationship with the earlier discussion on confidence intervals is clear. Using a negative estimate of $\hat{\beta}$ as our example, if we can reject $H_0: \beta=0$ with a confidence level of $C\%$ for a two-sided test, then this means that $0$ forms the upper bound of the $C\%$ confidence interval around the estimated value $\hat{\beta_k}$. 

It is also important to note that the definition of $\alpha_{\text{type I}}$ depends on the parametric sampling distribution, which varies by the type of test, holds asymptotically for DCM, and relies on specific assumptions. Consequently, the true (empirical) $\alpha_{\text{type I}}$ may differ from the calculated $\alpha_{\text{type I}}$. This difference is referred to as the size distortion of the statistical test. If the distortion is positive or ``liberal'', the empirical $\alpha_{\text{type I}}$ is larger than the calculated one, and this increases the likelihood of committing more Type I errors than expected. Conversely, if the distortion is negative or ``conservative'', the test makes fewer Type I errors than planned, but becomes more prone to Type II errors. Size distortion can be analyzed using Monte Carlo simulations, which replicate the resampling process and provide a useful tool for comparing the performance of different statistical tests.

The discussion above has shown the crucial role that $p$-values play in the decisions made by the analyst. While $p$-values obtained from LR and LM tests are unambiguous in their meaning (given the one-sided nature of these tests), the same is not true for Wald tests and thus $t$-ratios. An analyst needs to make an informed decision on whether a one-sided or two-sided test is used, i.e. how $\mathbf{H_1}$ is defined. If an analyst uses a two-sided test for a parameter with a clear sign assumption where a one-sided test would be more appropriate, then the $p$-value will be multiplied by a factor of $2$, reducing the likelihood of rejecting $\mathbf{H_0}$, and thus increasing the likelihood of a type II error.

A different perspective on the problem lies in considering the scientific value that emerges from the outcome of a test, the impact of significance on beliefs. Although it is usual practice to confer point rejections of the null (typically a p-value above 5\%) a larger scientific value, \cite{abadie2020statistical} remarks that the opposite is true instead. Given that often the prior of a rejection is larger than 50\%, a non-significant result (i.e. accepting the null) is more surprising, and consequently more informative than a significant result.

An additional point worth raising relates to the whether any corrections are needed to $p$-values as the estimation results of choice models imply the testing of multiple hypotheses at the same time (given that multiple parameters are estimated). In statistics, there is the notion that the probability of observing a rare event increases when conducting multiple tests, meaning that there would be increased risk of a type I error. Correction approaches have been put forward for this, including the Bonferroni correction \citep{bonferroni1936teoria} and various false discovery rate (FDR) correction approaches \citep[see e.g.][]{10.1111/j.2517-6161.1995.tb02031.x}, with the view to increasing $p$-values in the case of multiple tests. Clearly, estimation results from a choice model relate to testing multiple hypotheses on a dataset, most obviously in terms of significance tests for each individual parameter, and if for example 20 parameters all obtain a $p$-value of 0.05, then one could expect that for one of these, the $H_0$ of no effect is rejected incorrectly. What specific correction, if any, should be applied is however unclear given that choice models rely on joint estimation of all effects. Any correction of the $p$-values would also increase the risk of type II errors and encourage analysts to incorrectly remove parameters from their models, putting them at risk of inferior model performance and endogeneity bias. Analysts should at the very least however stay clear of statements that all parameters are different from zero, given the above point. For this reason, joint hypothesis tests, such as likelihood ratio tests, are generally better suited for assessing the relevance of groups of parameters in choice models than multiple individual $t$-tests, as they typically exhibit greater power for joint hypotheses and do not rely on ad-hoc multiple-testing corrections.

In the context of the earlier discussion about the trinity of tests, the relationship between statistical testing of the model structure in terms of the impact of adding or removing parameters, and tests for individual parameters becomes clear when talking about standard errors. Changes to the value of parameters with larger $t$-ratios (i.e. smaller standard errors relative to their estimates) will lead to larger changes in model fit. The removal of parameters with larger $t$-ratios will thus lead to a bigger drop in fit. At the same time, it is well known that as the sample size $N$ increases, the error decreases, with $\hat{\sigma}_{\hat{\beta}}$ being inversely proportional to $\sqrt{N}$. As a result, it can be easily understood that bigger changes in log-likelihood are commonly observed when adding or removing parameters in models with larger samples, and this justifies why measures such as the Bayesian Information Criterion (BIC) impose stricter criteria for including additional parameters with larger $N$. Let us take the case where an analyst compares two models, finding that the addition of a single parameter leads to an improvement in log-likelihood by 4 units, i.e. a likelihood ratio test value of $8$, which exceeds the $\chi^2_{1}$ 99\% critical value. This would in many papers be accompanied by an endorsement that this improvement is highly \emph{``significant''}. Using the above example of an improvement in fit by $4$ units, let us assume that this is for a dataset with $3$ alternatives, and an average probability of correct prediction of $60\%$ in the base model. In a dataset with $1,000$ observations, an improvement in log-likelihood by $4$ units would mean that the average probability of correct prediction has risen to $60.2\%$, for the sake of simplicity assuming equal benefit across observations. With $5,000$ observations, the improvement is even more negligible, at $60.05\%$. This calls for analysts to exercise restraint in describing improvements as large/significant on the basis of a likelihood improvement alone, and to consider also looking at goodness of fit measures where the penalty for additional parameters increases with the sample size (such as the BIC). In addition, there is merit in going further than model fit alone by also looking at the behavioural relevance of the additional parameters. This includes comparing different models in terms of whether key metrics such as willingness-to-pay and elasticities are in line with generally accepted ranges for these measures. %In this context, it is then also meaningful to contrast these measures across models not just in terms of their estimates, but also their statistical properties, in terms of confidence, for example. 

\subsection{Reporting of results}

The next consideration for an analyst is how significance levels should be reported, where we raise three key issues, namely interpretation, numerical precision and further use of results.

Remember that estimation gives values for $\hat{\beta_k}$ and $\hat{\sigma}_{k}$, i.e. estimates and standard errors. If an analyst includes these two outputs in the reporting of results, then this ensures that a reader is able to further process the results. The same applies if an analyst reports $\hat{\beta_k}$ and $t_{0,\hat{\beta}_k}$, i.e. the estimates and the $t$-ratios for parameters (against $\mathbf{H_0}$: $\beta_k = 0$), as a reader can calculate $\hat{\sigma}_{k}=\frac{\hat{\beta_k}}{t_{0,\hat{\beta}_k}}$. From this perspective, reporting either standard errors or $t$-ratios raises no further issues in theory. Some analysts may prefer $t$-ratios over standard errors as it allows for more direct insights into whether the findings can reject $\mathbf{H_0}: \beta_k = 0$ (and as they are dimension free), while others may prefer standard errors in order to compute confidence intervals or if they are interested in $t$-ratios against values other than $0$.

A practical consideration arises in terms of the numerical precision that is used in presenting the results. As standard errors are often an order of magnitude smaller than parameter estimates, the use of too low a number of digits will mean valuable information is lost in reporting results. Analysts should ensure to report at least two significant digits (i.e. not counting leading zeros). For $t$-ratios, it is common to report two digits, although the second decimal might be considered less useful as it relates to 1\% of a standard error. For parameter estimates, analysts may choose to report even more digits for special cases, while realising that reporting too large a number of digits claims spurious accuracy and confuses the reader. 

The situation with reporting becomes more complicated if an analyst reports $p$-values instead of $t$-ratios and standard errors, as is common practice in some disciplines, such as health, while being rare in others, such as transport. First, given the above points about one-sided \emph{vs} two-sided tests, if an analyst reports $p$-values only (i.e. without standard errors and/or $t$-ratios), then, in order to avoid misinterpretation, information must be provided on whether these relate to a one-sided or a two-sided test. Second, an issue with precision and further use of the results arises. Let us assume that we have $\mid t_{0,\hat{\beta}}\mid >10$, in which case, whether using a one-sided or a two-sided test, an analyst is likely going to report $p=0$ or $p<10^{-5}$ or, preferably, $p<.001$, the APA style option. Of course, the test clearly rejects $H_0: \beta=0$, but reporting $p$-values alone will prevent a reader from using the results to compute confidence intervals or to gain any insights into differences in uncertainty for two parameters which both have $p<10^{-5}$.

The issue is further compounded when an analyst relies on \emph{star} measures instead of $p$-values, e.g. using $^*$ for $90\%$ confidence, $^{**}$ for $95\%$ confidence and $^{***}$ for $99\%$ confidence. This again requires reporting whether one or two-sided tests were used, but issues with precision and further use of the results are even more severe than with $p$-values, given the arbitrary division of $p$-values into three categories. The further use of results for e.g. confidence intervals is impossible when significance levels are reported only with \emph{star} measures, even in those cases where with $p$-values, this would be possible still (i.e. if the reported $p$ is shown with enough digits).

A key motivation for including $p$-values is clearly to allow readers to quickly see significance levels without having to compare $t$-ratios to critical values. The same applies to \emph{star} measures, which are a somewhat more qualitative reporting tool as they simply split parameter estimates into four groups (i.e. $<90\%$ confidence, $90\%$ to $95\%$, $95\%$ to $99\%$, and $>99\%$). The use of these measures, accompanied by information on whether one-sided or two-sided tests are used, can thus be seen as providing additional information to readers. It is fine to report these measures alongside \underline{but not instead of} standard errors or $t$-ratios, but this of course goes against the motivation often given for the use of \emph{star} measures that it makes tables less cluttered and more readable.

\section{Empirical example}\label{sec:application}

The dataset used for our empirical example comes from a large-scale revealed preference (RP) survey conducted as part of the DECISIONS project carried out by the Choice Modelling Centre at the University of Leeds \citep[cf.][]{calastri2020we}. We specifically use the GPS tracking component of this survey to model the observed mode choice behaviour, with data processing and cleaning reported in \citet{tsoleridis2022deriving}.

For the present analysis, we focus on work trips, with a sample of 3,438 trips made by 358 individuals. For each trip, individuals travelled by one of six modes: car (47.6\%), bus (19.9\%), rail (8\%), taxi (1.5\%), cycling (3.9\%) or walking (19.1\%). The data covers a very wide range of trip distances, going from just under 100 metres to just under $100$ kilometres, with a mean distance of $10$ kilometres and a median distance of $5.9$ kilometres.

The attributes of the alternatives used in the models for the present paper include in-vehicle travel time, out-of-vehicle travel time, and travel cost. We used a linear-in-attributes specification of a Multinomial Logit (MNL) model, with mode specific in vehicle and out of vehicle travel time parameters. For this illustration, we rely on such a simple model, including without an explicit treatment of the repeated choice nature of the data. All models were estimated using Apollo \citep{hess_palma_apollo}\footnote{The data and code are available via \url{www.stephanehess.me.uk/papers/significance.zip}.}. In addition to the standard estimation (and computation of classical and robust standard errors), we also conducted bootstrap estimation using $400$ samples taken at the level of individual travellers (as opposed to individual trips). 

The results of our analysis are shown in Table \ref{tab:example}. We follow the policy of showing at least 2 significant digits (i.e. not counting leading zeros) for estimates, standard errors and confidence limits, which in our case is possible by showing four digits after the decimal point throughout, and 5 in one case. For the latter two, we use scientific notation giving the larger number of leading zeros for some parameters. For p-values, we show three digits after the decimal point, adopting the APA convention of $p<.001$ where this applies. 

Looking at the estimates ($\hat{\beta}_k$), we note that, with car as the base, the alternative specific constants (ASC) for all other modes except walking are negative. In line with expectations, the in vehicle travel time, out of vehicle travel time, and travel cost coefficients are all negative, with higher out of vehicle than in vehicle time sensitivites. 

\begin{sidewaystable}[htbp]
  \centering\small\setlength{\tabcolsep}{1.5pt}
    \caption{Results on DECISIONS mode choice data}
  \resizebox{\textwidth}{!}{%
        \begin{tabular}{r:c:c:c:c:c:c:c:c:c:c:c:c:c:c}
            \toprule
        & $\delta_{bus}$ & $\delta_{rail}$ & $\delta_{taxi}$ & $\delta_{cycling}$ & $\delta_{walking}$ & $\beta_{car,tt}$ & $\beta_{bus,tt}$ & $\beta_{rail,tt}$ & $\beta_{taxi,tt}$ & $\beta_{cycling,tt}$ & $\beta_{walking,tt}$ & $\beta_{bus,ovt}$ & $\beta_{rail,ovt}$ & $\beta_{tc}$ \\
          \midrule
    $\hat{\beta}_k$ & -1.9464 & -0.4801 & -4.9270 & -3.2225 & 0.2193 & -0.1236 & -0.0439 & -0.0433 & -0.0142 & -0.0745 & -0.1107 & -0.0792 & -0.1088 & -0.2230 \\
    $\hat{\sigma}_k$ & 0.1507 & 0.3119 & 0.2970 & 0.2140 & 0.2447 & 0.0096 & 0.0051 & 0.0097 & 0.0165 & 0.0088 & 0.0093 & 0.0094 & 0.0105 & 0.0164 \\
    $\sigma_{robust,k}$ & 0.4977 & 0.5793 & 0.8850 & 0.5122 & 0.4431 & 0.0256 & 0.0093 & 0.0235 & 0.0509 & 0.0166 & 0.0200 & 0.0565 & 0.0213 & 0.0370 \\
    $\sigma_{bootstrap,k}$ & 0.4772 & 0.6067 & 1.0139 & 0.5307 & 0.4451 & 0.0250 & 0.0094 & 0.0248 & 0.0618 & 0.0169 & 0.0208 & 0.0543 & 0.0227 & 0.0394 \\
          \midrule
    $t_{0,\beta_k} \text{ (classical)}$ & -12.91 & -1.54 & -16.59 & -15.06 & 0.90  & -12.83 & -8.58 & -4.45 & -0.86 & -8.48 & -11.88 & -8.44 & -10.40 & -13.61 \\
    $p_{t_{0,\beta_k},classical,1-sided}$ & $<.001$ *** & 0.062 * & $<.001$ *** & $<.001$ *** & 0.185 & $<.001$ *** & $<.001$ *** & $<.001$ *** & 0.195 & $<.001$ *** & $<.001$ *** & $<.001$ *** & $<.001$ *** & $<.001$ *** \\
    $p_{t_{0,\beta_k},classical,2-sided}$ & $<.001$ *** & 0.124 & $<.001$ *** & $<.001$ *** & 0.370 & $<.001$ *** & $<.001$ *** & $<.001$ *** & 0.390 & $<.001$ *** & $<.001$ *** & $<.001$ *** & $<.001$ *** & $<.001$ *** \\
    $t_{0,\beta_k} \text{ (robust)}$ & -3.91 & -0.83 & -5.57 & -6.29 & 0.49  & -4.83 & -4.70 & -1.84 & -0.28 & -4.49 & -5.53 & -1.40 & -5.11 & -6.03 \\
    $p_{t_{0,\beta_k},robust,1-sided}$ & $<.001$ *** & 0.204 & $<.001$ *** & $<.001$ *** & 0.310 & $<.001$ *** & $<.001$ *** & 0.033 ** & 0.390 & $<.001$ *** & $<.001$ *** & 0.080 * & $<.001$ *** & $<.001$ *** \\
    $p_{t_{0,\beta_k},robust,2-sided}$ & $<.001$ *** & 0.407 & $<.001$ *** & $<.001$ *** & 0.621 & $<.001$ *** & $<.001$ *** & 0.065 * & 0.780 & $<.001$ *** & $<.001$ *** & 0.161 & $<.001$ *** & $<.001$ *** \\
    $t_{0,\beta_k} \text{ (bootstrap)}$ & -4.08 & -0.79 & -4.86 & -6.07 & 0.49  & -4.95 & -4.69 & -1.75 & -0.23 & -4.40 & -5.33 & -1.46 & -4.79 & -5.66 \\
    $p_{t_{0,\beta_k},bootstrap,asymptotic 1-sided}$ & $<.001$ *** & 0.214 & $<.001$ *** & $<.001$ *** & 0.311 & $<.001$ *** & $<.001$ *** & 0.040 ** & 0.409 & $<.001$ *** & $<.001$ *** & 0.072 * & $<.001$ *** & $<.001$ *** \\
    $p_{t_{0,\beta_k},bootstrap,asymptotic 2-sided}$ & $<.001$ *** & 0.429 & $<.001$ *** & $<.001$ *** & 0.622 & $<.001$ *** & $<.001$ *** & 0.080 * & 0.819 & $<.001$ *** & $<.001$ *** & 0.144 & $<.001$ *** & $<.001$ *** \\
    $p_{t_{0,\beta_k},bootstrap,empirical 1-sided}$ & $<.001$ *** & 0.232 & $<.001$ *** & $<.001$ *** & 0.263 & $<.001$ *** & $<.001$ *** & 0.043 ** & 0.374 & $<.001$ *** & $<.001$ *** & 0.020 ** & $<.001$ *** & $<.001$ *** \\
    $p_{LR}$ & $<.001$ *** & 0.126 & $<.001$ *** & $<.001$ *** & 0.368 & $<.001$ *** & $<.001$ *** & $<.001$ *** & 0.384 & $<.001$ *** & $<.001$ *** & $<.001$ *** & $<.001$ *** & $<.001$ *** \\
    $p_{LM}$ & $<.001$ *** & 0.124 & $<.001$ *** & $<.001$ *** & 0.370 & $<.001$ *** & $<.001$ *** & $<.001$ *** & 0.389 & $<.001$ *** & $<.001$ *** & $<.001$ *** & $<.001$ *** & $<.001$ *** \\
          \midrule
    $L_{\beta_k,classical}$ & -2.2419 & -1.0915 & -5.5092 & -3.6418 & -0.2602 & -0.1425 & -0.0539 & -0.0624 & -0.0465 & -0.0917 & -0.1290 & -0.0976 & -0.1293 & -0.2551 \\
    $U_{\beta_k,classical}$ & -1.6510 & 0.1313 & -4.3448 & -2.8032 & 0.6988 & -0.1047 & -0.0339 & -0.0242 & 0.0181 & -0.0572 & -0.0924 & -0.0608 & -0.0883 & -0.1909 \\
    $L_{\beta_k,robust}$ & -2.9219 & -1.6156 & -6.6616 & -4.2264 & -0.6491 & -0.1738 & -0.0622 & -0.0893 & -0.1139 & -0.1070 & -0.1499 & -0.1900 & -0.1506 & -0.2955 \\
    $U_{\beta_k,robust}$ & -0.9709 & 0.6553 & -3.1925 & -2.2186 & 1.0877 & -0.0734 & -0.0256 & 0.0027 & 0.0855 & -0.0419 & -0.0715 & 0.0315 & -0.0671 & -0.1506 \\
    $L_{\beta_k,bootstrap}$ & -2.7926 & -1.6724 & -7.0079 & -4.3318 & -0.6158 & -0.1774 & -0.0662 & -0.0956 & -0.1606 & -0.1141 & -0.1578 & -0.1800 & -0.1554 & -0.3129 \\
    $U_{\beta_k,bootstrap}$ & -0.9863 & 0.7287 & -3.0426 & -2.2758 & 1.1413 & -0.0804 & -0.0285 & 0.0042 & 0.0766 & -0.0457 & -0.0741 & -0.00055 & -0.0742 & -0.1552 \\
    $L_{\beta_k,HPD}$ & -2.7472 & -1.4885 & -6.8906 & -4.2116 & -0.5746 & -0.1689 & -0.0639 & -0.0916 & -0.1464 & -0.1091 & -0.1608 & -0.1744 & -0.1577 & -0.3040 \\
    $U_{\beta_k,HPD}$ & -0.9574 & 0.8339 & -3.0243 & -2.2092 & 1.1497 & -0.0747 & -0.0265 & 0.0080 & 0.0795 & -0.0412 & -0.0789 & 0.0018 & -0.0748 & -0.1501 \\
    width robust \emph{vs} classical & 3.30  & 1.86  & 2.98  & 2.39  & 1.81  & 2.66  & 1.83  & 2.41  & 3.09  & 1.89  & 2.15  & 6.02  & 2.03  & 2.26 \\
    width bootstrap \emph{vs} robust & 0.93  & 1.06  & 1.14  & 1.02  & 1.01  & 0.97  & 1.03  & 1.08  & 1.19  & 1.05  & 1.07  & 0.81  & 0.97  & 1.09 \\
    width HPD \emph{vs} empirical bootstrap & 0.99  & 0.97  & 0.98  & 0.97  & 0.98  & 0.97  & 0.99  & 1.00  & 0.95  & 0.99  & 0.98  & 0.98  & 1.02  & 0.98 \\
    empirical bootstrap asymmetry index & 0.06  & 0.01  & -0.05 & -0.08 & 0.05  & -0.11 & -0.18 & -0.05 & -0.23 & -0.16 & -0.13 & -0.12 & -0.15 & -0.14 \\
    HPD asymmetry index & 0.11  & 0.13  & -0.02 & 0.01  & 0.08  & 0.04  & -0.07 & 0.03  & -0.17 & -0.02 & -0.22 & -0.08 & -0.18 & -0.05 \\
 \bottomrule
       \end{tabular}%
    }
  \label{tab:example}%
  \flushleft{$^{***}: p\leq 0.01$; $^{**}: p\leq 0.05$; $^{*}: p\leq 0.1$}
\end{sidewaystable}%

We first focus on the standard errors. For all $14$ model parameters, the classical standard errors ($\hat{\sigma}_k$) are substantially smaller than the robust standard errors ($\sigma_{robust,k}$), which is in line with general findings, and is likely in part related to the lack of an explicit treatment of the repeated choice nature of the data in our model. The bootstrap standard errors ($\sigma_{bootstrap,k}$) are closer to the robust standard errors, which they exceed for $10$ out of $14$ parameters.

We first look at the trinity of statistical tests for the individual parameters. We present the $p$-values for one-sided and two-sided asymptotic $t$-ratios using classical, robust and bootstrap standard errors. In addition, we report the $p$-value for a one-sided empirical bootstrap $t$-ratio, which is the share of bootstrap estimates that are the other side of zero from $\hat{\beta}_k$. Finally, we also show the $p$-values for the likelihood ratio (LR) and Lagrange multiplier (LM) tests. For $9$ parameters ($\delta_{bus}$, $\delta_{taxi}$, $\delta_{cycling}$, $\beta_{car,tt}$, $\beta_{bus,tt}$, $\beta_{cycling,tt}$, $\beta_{walking,tt}$, $\beta_{rail,ovt}$, $\beta_{tc}$), all $9$ of the tests reject $\mathbf{H_0}$: $\beta_k = 0$ above the 99\% level of confidence, i.e. with $p<0.01$.

We now discuss the findings for the remaining five coefficients in turn. For $\delta_{rail}$, only the classical one-sided $t$-ratio rejects the null hypothesis of no effect at any usually accepted level of confidence ($94\%$ in this case). For $\delta_{walking}$, none of the tests reject the null hypothesis of no effect. However, an analyst would be ill advised to remove these ASCs from the model as the null hypothesis that the net effect of omitted variables is zero has little credibility and the ASCs also ensure the recovery of the market shares in the data. For $\beta_{rail,tt}$, the classical $t$-ratios (one-sided and two-sided) reject $\mathbf{H_0}$ above the $99\%$ level of confidence, as do the LR and LM tests. For the robust and bootstrap tests, the significance level changes, where a one-sided test now only allows us to reject $\mathbf{H_0}$ at the $96\%-97\%$ level of confidence. This is an illustration of the potential for different findings using different tests, with such differences being more likely (and potentially larger) in smaller samples, leading to the possibility of one test rejecting $\mathbf{H_0}$ while another fails to do so. The in vehicle time parameter for taxi ($\beta_{taxi,tt}$) is an example of a case where behavioural/policy relevance trumps statistical significance. All tests fail to reject $\mathbf{H_0}$ of no effect, yet any transport modeller would be ill advised to remove such a policy relevant variable from the model, and should acknowledge that the low choice rate for taxi likely plays a role. Finally, $\beta_{bus,ovt}$ presents a case similar to $\beta_{rail,tt}$. The classical $t$-ratios (one-sided and two-sided) reject $\mathbf{H_0}$ above the $99\%$ level of confidence, as do the LR and LM tests. For the robust and bootstrap tests, the significance level changes, where a one-sided test now only allows us to reject $\mathbf{H_0}$ at the $92\%$ (robust), $93\%$ (bootstrap) and $98\%$ (empirical bootstrap) level of confidence.

We finally look at confidence limits (denoted as $L$ and $U$ for the lower and upper limits of a $95\%$ confidence interval). The differences in width for the classical and robust confidence intervals are in line with the earlier discussions about classical \emph{vs} robust standard errors. An additional insight arises when we compare the width of confidence intervals across parameters. Taking the example of $\beta_{car,tt}$, $\beta_{bus,tt}$ and $\beta_{rail,tt}$, all three parameters reject $\mathbf{H_0}$ of no effect above the $99\%$ level of confidence, but obtain very different confidence intervals. Indeed, for $\beta_{car,tt}$, a $95\%$ CI stretches $15.27\%$ either side of $\hat{\beta}_k$, while this is $22.85\%$ for $\beta_{bus,tt}$, and $44.09\%$ for $\beta_{rail,tt}$. This is a striking difference and highlights the distinction between \emph{significance} and \emph{precision} of an estimate. Although we can easily reject $\mathbf{H_0}$ for all three parameters, the findings for confidence intervals could have implications for policy. 

We finally return to the issue raised early on about asymptotic normality. The use of the formula in Equation (\ref{eq:CI}) uses the more stringent property of normality, which may not be appropriate outside the immediate neighbourhood of the optimum. To investigate this issue, we contrast the asymptotic confidence intervals with those obtained using bootstrapping. Of course, the bootstrap and robust confidence intervals are wider than those obtained using the classical standard errors, in line with the difference in the standard errors. However, we also note that the bootstrap confidence intervals are not necessarily symmetric around the MLE. We calculate an asymmetry index as $\frac{(U-M)-(M-L)}{U-L}$, where $0$ would indicate symmetry. For the ASCs, we see negative skew for the CIs for $\delta_{taxi}$ and $\delta_{cycling}$, with positive skew for the remaining constants. More importantly, the CIs are negatively skewed for all $9$ parameters relating to travel time and travel costs. Comparing the empirical bootstrap CIs with the HDP intervals, we see that the latter are narrower for $13$ out of $14$ parameters, in line with earlier discussions. A majority of the CIs retain a negative skew, but this now changes to positive for $\delta_{cycling}$, $\beta_{car,tt}$ and $\beta_{rail,tt}$. 

\section{Discussion and conclusions}\label{sec:conclusions}

This note has looked at the issues of the interpretation and reporting of measures of statistical confidence in the context of choice model estimation. We have provide an in-depth overview of the reason for parameter uncertainty, its computation, and its use in confidence intervals and hypothesis testing. The misuse of the notion of ``statistical significance'' has been a long-standing concern for statisticians in particular and science more widely. The authoritative paper by \citet{Wasserstein_et_al} goes as far as concluding \emph{``that it is time to stop using the term `statistically significant' entirely. Nor should variants such as `significantly different', `p<0.05', and `nonsignificant' survive, whether expressed in words, by asterisks in a table, or in some other way.''} Less controversially, they say that \emph{``[analysts should not] believe that an association or effect exists just because it was statistically significant [or] that an association or effect is absent just because it was not statistically significant.''}

We share the view that excessive attention is paid to the statistical properties of findings, often at the expense of the behavioural and policy implications. At the same time, we do not foresee a situation where choice modellers will abandon the notion completely, at least any time soon, and this note thus provides some guidance for applied work. We believe that:

\begin{itemize}
    \item Greater care is required with language, noting that the commonly used notion that a parameter is ``statistically significant'' is incorrect, as the significance relates to the probability of a type I error. The correct wording is that we can reject the null hypothesis of no effect at a given level of confidence or significance.
    \item Analysts should move away from the notion of $95\%$ confidence (i.e. $p=0.05$) being a hard rule, recognising that this level is often easily obtained with large datasets, that lower thresholds may be acceptable for smaller datasets, and that a parameter can provide important behavioural insights even if it has a lower level of significance. Failure to achieve sufficiently small $p$-values  may simply indicate that not enough data has been collected. On the other hand, a $p$-value of 0.05 implies that the impact of the concerned variable has approximate confidence limits of +/- $100\%$.
    \item Analysts should look further than statistical significance alone. In a health context, it is common to talk about \emph{``clinical importance''}, which measures whether a treatment has a genuine and noticeable effect on health outcomes. In choice modelling more broadly, analysts may wish to consider the \emph{``behavioural importance''} of a parameter, i.e. whether it changes predictions (cf. discussions in Section \ref{sec:model_comparison} and the \emph{``policy importance''}, i.e. whether a finding has a sizeable impact on the outcome of any process using the results). 
    \item Analysts should take care to report measures of parameter uncertainty in a numerically precise way. This applies to standard errors, $t$-ratios, and $p$-values.
    \item Analysts need to properly document their test findings, by clarifying whether these are for one-sided or two-sided tests. 
    \item If analysts wish to highlight \emph{significant} parameters using \emph{star} measures, this should only be done if standard errors or $t$-ratios are presented alongside them. With $p$-values, analysts should similarly  present these measures not in isolation but alongside standard errors or $t$-ratios.
    \item A distinction should be made between parameter significance and precision, noting that two parameters that both pass a significance test above the $99\%$ level can have vastly different confidence intervals. In practical work, precision may matter much more than statistical significance, and analysts may want to ensure that an effect is behaviourally meaningful and important across the entire width of say a $95\%$ confidence interval.
    \item The property of asymptotic normality needs to be better understood, and analysts should recognise that computing confidence intervals in particular may be very inaccurate by assuming that normality holds several standard errors away from the MLE. Bootstrapped confidence intervals may then be preferable.
\end{itemize}

An issue that deserves some discussion at the end is that of researcher incentives and the related problem of $p$-hacking, which essentially implies coming up with a specification that gives an analyst the \emph{right} $t$-ratio for the effect they're interested in. Choice modelling is largely an empirical field, and it is noticeable how in some topic areas, a lot of emphasis is placed on a priori registration of hypotheses, while in other areas, a more data-driven approach is used. While the latter is often described as fishing for evidence (or significance), it should also be clear that the former carries risks too, namely in terms of findings being biased by effects that are unaccounted for. Furthermore, analysts wary of the straitjacket of pre-registration may be tempted to develop a near-complete list of possible hypotheses, which defeats the entire point. However, such bad practice is unlikely to change without better editorial practice and improved teaching of statistics, not just for authors but also for reviewers who continuously call for tables of results filled with stars and only \emph{significant} parameters. 

An additional caveat relates to the disconnect between applied choice modelling practice and inference theory. The latter is based on the idea that the model is fixed. In practice, we conduct tests on individual parameters for one model at a time, and these tests only tell us something subject to the hypothesis that this is the true model. A different conclusion could be reached with a different model. Different $p$-values will be obtained for the same parameter in different models, which relates to the earlier point that the wrongly specified model has not only wrong estimates but also wrong statistical significance properties.

Finally, and as an alternative to frequentist significance testing, analysts could consider the Bayesian approach to parameter uncertainty, where prior knowledge is updated in response to evidence provided by data, and statistical inference is based upon decision-making theory. Through the concept of posterior probabilities, Bayesian econometrics offers a probabilistic representation of not only parameters but also hypotheses. For instance, Bayes estimates are the whole posterior distribution of the parameters from which analysts can derive the exact probability of (credible) intervals containing the true parameter. Bayes’ rule can also be used to derive measures of relative evidence across competing models, including posterior probabilities of  hypotheses. For example, and unlike $p$-values, Bayes factors provide a data-supported measure of the odds in favour of the null hypothesis over the alternative. Even though Bayesian hypothesis testing is more intuitive and departs from frequentist concepts, there is an open discussion among Bayesian practitioners about the flaws in null hypothesis significance testing, which goes beyond the scope of this note, and we refer the reader to \citet{mcelreath2020statistical}.

\section*{Acknowledgements}

Stephane Hess acknowledges the support of the European Research Council through the advanced grant 101020940-SYNERGY. Angelo Guevara acknowledges the support by ANID FONDECYT 1231584 and ANID PIA/PUENTE AFB220003. The authors would like to thank David Bunch for feedback on earlier work. We are also grateful to two anonymous referees whose comments substantially improved the paper.

\bibliographystyle{elsarticle-harv}
\bibliography{refs,ref_library}

@article{Imbens_2021,
Author = {Imbens, Guido W.},
Title = {Statistical Significance, p-Values, and the Reporting of Uncertainty},
Journal = {Journal of Economic Perspectives},
Volume = {35},
Number = {3},
Year = {2021},
Month = {August},
Pages = {157–74}}

@misc{hess2025holeprofilelikelihoodapproach,
      title={Get me out of this hole: a profile likelihood approach to identifying and avoiding inferior local optima in choice models}, 
      author={Stephane Hess and David Bunch and Andrew Daly},
      year={2025},
      eprint={2506.02722},
      archivePrefix={arXiv},
      primaryClass={econ.EM},
      url={https://arxiv.org/abs/2506.02722}, 
}

@book{bonferroni1936teoria,
  title={Teoria statistica delle classi e calcolo delle probabilit{\`a}},
  author={Bonferroni, C.E.},
  series={Pubblicazioni del R. Istituto superiore di scienze economiche e commerciali di Firenze},
  url={https://books.google.es/books?id=3CY-HQAACAAJ},
  year={1936},
  publisher={Seeber}
}

@article{10.1111/j.2517-6161.1995.tb02031.x,
    author = {Benjamini, Yoav and Hochberg, Yosef},
    title = {Controlling the False Discovery Rate: A Practical and Powerful Approach to Multiple Testing},
    journal = {Journal of the Royal Statistical Society: Series B (Methodological)},
    volume = {57},
    number = {1},
    pages = {289-300},
    year = {2018},
    month = {12},
    issn = {0035-9246},
    doi = {10.1111/j.2517-6161.1995.tb02031.x}
}

@article{DALY2023103828,
title = {Estimating willingness-to-pay from discrete choice models: Setting the record straight},
journal = {Transportation Research Part A: Policy and Practice},
volume = {176},
pages = {103828},
year = {2023},
issn = {0965-8564},
doi = {https://doi.org/10.1016/j.tra.2023.103828},
author = {Andrew Daly and Stephane Hess and {\relax J.\space{de D}} Ort{\'{u}}zar}}

@book{gelman2013bayesian,
  title={Bayesian Data Analysis, Third Edition},
  author={Gelman, A. and Carlin, J.B. and Stern, H.S. and Dunson, D.B. and Vehtari, A. and Rubin, D.B.},
  isbn={9781439840955},
  lccn={2013039507},
  series={Chapman \& Hall/CRC Texts in Statistical Science},
  url={https://books.google.co.uk/books?id=ZXL6AQAAQBAJ},
  year={2013},
  publisher={Taylor \& Francis}
}

@Article{coda,
    title = {CODA: Convergence Diagnosis and Output Analysis for MCMC},
    author = {Martyn Plummer and Nicky Best and Kate Cowles and Karen
      Vines},
    journal = {R News},
    year = {2006},
    volume = {6},
    number = {1},
    pages = {7--11},
    url = {https://journal.r-project.org/archive/},
  }

@article{Gudicha2016StatisticalPO,
  title={Statistical power of likelihood ratio and Wald tests in latent class models with covariates},
  author={Dereje W. Gudicha and Verena D. Schmittmann and Jeroen K. Vermunt},
  journal={Behavior Research Methods},
  year={2016},
  volume={49},
  pages={1824 - 1837},
  url={https://api.semanticscholar.org/CorpusID:11972846}
}

@Article{pub:81584,
 title                = {Sample Size Requirements for Discrete-Choice Experiments in Healthcare: a Practical Guide},
 author               = {de Bekker-Grob, Esther and Donkers, Bas and Jonker, Marcel F. and Stolk, Elly},
 journal              = {The Patient: patient-centered outcomes research},
 volume               = {8},
 number               = {5},
 pages                = {373--384},
 year                 = 2015,
 month                = oct,
 doi                  = {10.1007/s40271-015-0118-z}
}

@Article{406,
  Title                    = {Confidence interval to bound the value of time},
  Author                   = {P. Armstrong and R. A. Garrido and {\relax J.\space{de D}} Ort{\'{u}}zar},
  Journal                  = {Transportation Research Part E},
  Year                     = {2001},
  Number                   = {1},
  Pages                    = {143--161},
  Volume                   = {37}
}

@Article{1,
  Title                    = {{Simulation estimation of mixed discrete choice models using randomized and scrambled Halton sequences}},
  Author                   = {C. R. Bhat},
  Journal                  = {Transportation Research Part B},
  Year                     = {2003},
  Number                   = {9},
  Pages                    = {837--855},
  Volume                   = {37}
}

@Book{100,
  Title                    = {The acceptance of modal innovation : The case of Swissmetro},
  Author                   = {M. Bierlaire and K. W. Axhausen and G. Abay},
  Publisher                = {Proceedings of the $1^{st}$ Swiss Transport Research Conference},
  Year                     = {2001},

  Address                  = {Monte Verit{\`a}, Ascona}
}

@Article{957,
  Title                    = {Masking identification of discrete choice models under simulation methods},
  Author                   = {L. Chiou and J. Walker},
  Journal                  = {Journal of Econometrics},
  Year                     = {2007},
  Number                   = {2},
  Pages                    = {683--703},
  Volume                   = {141}
}

@InCollection{760,
  Title                    = {{Eliminating bias due to the repeated measurements problem in SP data}},
  Author                   = {C. Cirillo and K. Lindveld and A. Daly},
  Booktitle                = {Stated Preference Modelling Techniques: PTRC Perspectives 4},
  Publisher                = {PTRC Education and Research Services Ltd},
  Year                     = {2000},

  Address                  = {London},
  Editor                   = {{\relax J.\space{de D}} Ort{\'{u}}zar}
}

@Article{999,
  Title                    = {On approximating the statistical properties of elasticities},
  Author                   = {I. Krinsky and A. Robb},
  Journal                  = {Review of Economics and Statistics},
  Year                     = {1986},
  Pages                    = {715--719},
  Volume                   = {68}
}

@InCollection{2,
  Title                    = {New developments in uniform pseudo-random number and vector generation},
  Author                   = {H. Niederreiter},
  Booktitle                = {Monte Carlo and Quasi-Monte Carlo Methods in Scientific Computing},
  Publisher                = {Springer},
  Year                     = {1995},

  Address                  = {New York},
  Editor                   = {Niederreiter, H. and Shiue, J-S.},
  Pages                    = {94--112}
}

@Book{799,
  Title                    = {Modelling Transport},
  Author                   = {{\relax J.\space{de D}} Ort{\'{u}}zar and L. G. Willumsen},
  Publisher                = {John Wiley and Sons},
  Year                     = {2011},

  Address                  = {Chichester},
  Edition                  = {$4^\text{th}$}
}

@Article{Bierlaire2010,
  author    = {Bierlaire, Michel and Th{\'e}mans, Micha{\"e}l and Zufferey, Nicolas},
  title     = {A heuristic for nonlinear global optimization},
  journal   = {INFORMS Journal on Computing},
  year      = {2010},
  volume    = {22},
  number    = {1},
  pages     = {59--70},
  publisher = {INFORMS},
}

@Article{Revelt1998,
  author    = {Revelt, David and Train, Kenneth},
  title     = {Mixed logit with repeated choices: households' choices of appliance efficiency level},
  journal   = {Review of economics and statistics},
  year      = {1998},
  volume    = {80},
  number    = {4},
  pages     = {647--657},
  publisher = {MIT Press},
}

@Article{hess_palma_apollo,
  author   = {Stephane Hess and David Palma},
  journal  = {Journal of Choice Modelling},
  title    = {Apollo: A flexible, powerful and customisable freeware package for choice model estimation and application},
  year     = {2019},
  issn     = {1755-5345},
  pages    = {100170},
  volume   = {32},
  doi      = {https://doi.org/10.1016/j.jocm.2019.100170}
}

@article{King_Roberts_2015, title={How Robust Standard Errors Expose Methodological Problems They Do Not Fix, and What to Do About It}, volume={23}, DOI={10.1093/pan/mpu015}, number={2}, journal={Political Analysis}, author={King, Gary and Roberts, Margaret E.}, year={2015}, pages={159–179}}

@book{LehmannCasella1998,
  author = {Lehmann, E. L. and Casella, G.},
  title = {Theory of Point Estimation},
  year = {1998},
  publisher = {Springer},
  edition = {Second},
  address = {New York},
}

@book{Ziliak,
 ISBN = {9780472070077},
 author = {Stephen T. Ziliak and Deirdre N. McCloskey},
 publisher = {University of Michigan Press},
 title = {The Cult of Statistical Significance: How the Standard Error Costs Us Jobs, Justice, and Lives},
 urldate = {2025-12-16},
 year = {2008}
}

@book{alma990024527500403126,
author = {Cameron, A. Colin and Trivedi, P. K.},
address = {Cambridge ;},
booktitle = {Microeconometrics : methods and applications},
isbn = {9780521848053},
language = {eng},
lccn = {2004022273},
publisher = {Cambridge University Press},
title = {Microeconometrics : methods and applications },
year = {2005},
}

@article{calastri2020we,
  title={We want it all: experiences from a survey seeking to capture social network structures, lifetime events and short-term travel and activity planning},
  author={Calastri, Chiara and dit Sourd, Romain Crastes and Hess, Stephane},
  journal={Transportation},
  volume={47},
  number={1},
  pages={175--201},
  year={2020},
  publisher={Springer}
}

@article{tsoleridis2022deriving,
  title={Deriving transport appraisal values from emerging revealed preference data},
  author={Tsoleridis, Panagiotis and Choudhury, Charisma F and Hess, Stephane},
  journal={Transportation Research Part A: Policy and Practice},
  volume={165},
  pages={225--245},
  year={2022},
  publisher={Elsevier}
}

@article{Parady_2023,
   title={Size matters: the use and misuse of statistical significance in discrete choice models in the transportation academic literature},
   volume={51},
   ISSN={1572-9435},
   DOI={10.1007/s11116-023-10423-y},
   number={6},
   journal={Transportation},
   publisher={Springer Science and Business Media LLC},
   author={Parady, Giancarlos and Axhausen, Kay W.},
   year={2024},
   month=sep, pages={2393–2425} }

@article{ASA,
author = {Ronald L. Wasserstein and Nicole A. Lazar},
title = {The ASA Statement on p-Values: Context, Process, and Purpose},
journal = {The American Statistician},
volume = {70},
number = {2},
pages = {129-133},
year  = {2016},
publisher = {Taylor & Francis},
doi = {10.1080/00031305.2016.1154108}
}

@INCOLLECTION{RePEc:nbr:nberch:10206,
title = {Estimation and Inference in Nonlinear Structural Models},
author = {Berndt, Ernst R. and Hall, Bronwyn and Hall, Robert and Hausman, Jerry},
year = {1974},
pages = {653-665},
booktitle = {Annals of Economic and Social Measurement, Volume 3, number 4},
publisher = {National Bureau of Economic Research, Inc},
url = {https://EconPapers.repec.org/RePEc:nbr:nberch:10206}
}

@article{Amrhein_et_al,
author = {Valentin Amrhein and Sander Greenland  and Blake McShane},
title = {Redefine statistical significance},
journal = {Nature Human Behaviour},
volume = 2,
pages = {6–10},
year = {2018},
doi = {10.1038/s41562-017-0189-z}}

@article{Benjamin_et_al,
author = {Benjamin, Daniel and Berger, James and Johannesson, Magnus and Nosek, Brian and Wagenmakers, E.-J and Berk, Richard and Bollen, Kenneth and Brembs, Björn and Brown, Lawrence and Camerer, Colin and Cesarini, David and Chambers, Christopher and Clyde, Merlise and Cook, Thomas and De Boeck, Paul and Dienes, Zoltan and Dreber, Anna and Easwaran, Kenny and Efferson, Charles and Johnson, Valen},
year = {2017},
month = {09},
pages = {},
title = {Redefine Statistical Significance},
volume = {2},
journal = {Nature Human Behaviour},
doi = {10.1038/s41562-017-0189-z}
}

@article{Wasserstein_et_al,
author = {Ronald L. Wasserstein and Allen L. Schirm and Nicole A. Lazar},
title = {Moving to a World Beyond ``p < 0.05''},
journal = {The American Statistician},
volume = {73},
number = {sup1},
pages = {1-19},
year  = {2019},
publisher = {Taylor & Francis},
doi = {10.1080/00031305.2019.1583913}
}

@Article{Daly2012a,
  author  = {Daly, A.J. and Hess, S. and de Jong, G.},
  title   = {Calculating errors for measures derived from choice modelling estimates},
  journal = {Transportation Research Part B},
  year    = {2012},
  volume  = {46},
  number  = {2},
  pages   = {333--341},
}

@incollection{guevara2024endogeneity,
  title={Endogeneity in discrete choice models},
  author={Guevara, C Angelo},
  booktitle={Handbook of Choice Modelling},
  pages={668--692},
  year={2024},
  publisher={Edward Elgar Publishing}
}

@article{guevara2018overidentification,
  title={Overidentification tests for the exogeneity of instruments in discrete choice models},
  author={Guevara, C Angelo},
  journal={Transportation Research Part B: Methodological},
  volume={114},
  pages={241--253},
  year={2018},
  publisher={Elsevier}
}

@article{ben1986akaike,
  title={The Akaike likelihood ratio index},
  author={Ben-Akiva, Moshe and Swait, Joffre},
  journal={Transportation Science},
  volume={20},
  number={2},
  pages={133--136},
  year={1986},
  publisher={INFORMS}
}

@ARTICLE{RePEc:eee:eejocm:v:21:y:2016:i:c:p:60-65,
title = {Discrete choice models' $\rho^2$: A reintroduction to an old friend},
author = {Mokhtarian, Patricia},
year = {2016},
journal = {Journal of choice modelling},
volume = {21},
number = {C},
pages = {60-65},
url = {https://EconPapers.repec.org/RePEc:eee:eejocm:v:21:y:2016:i:c:p:60-65}
}

@techreport{domencich1975urban,
  title={Urban travel demand-a behavioral analysis},
  author={Domencich, Thomas A and McFadden, Daniel},
  year={1975},
  institution={Northwestern University},
  publisher={North - Holland, 1975}
}

@article{abadie2020statistical,
  title={Statistical nonsignificance in empirical economics},
  author={Abadie, Alberto},
  journal={American Economic Review: Insights},
  volume={2},
  number={2},
  pages={193--208},
  year={2020},
  publisher={American Economic Association 2014 Broadway, Suite 305, Nashville, TN 37203}
}

@article{hausman1984specification,
  title={Specification tests for the multinomial logit model},
  author={Hausman, Jerry and McFadden, Daniel},
  journal={Econometrica: Journal of the econometric society},
  pages={1219--1240},
  year={1984},
  publisher={JSTOR}
}

@book{gelman2020regression,
  title={Regression and Other Stories},
  author={Gelman, Andrew and Hill, Jennifer and Vehtari, Aki},
  year={2020},
  publisher={Cambridge University Press},
  address={Cambridge, UK},
  isbn={978-1107023987},
  url={https://avehtari.github.io/ROS-Examples/}
}

@Article{repec:eee:transb:v:58:y:2013:i:c:p:199-214,
journal={Transportation Research Part B: Methodological},
author={Bliemer, Michiel C.J. and Rose, John M.},
title={Confidence intervals of willingness-to-pay for random coefficient logit models},
year={2013},
month={None},
pages={199-214},
volume={58},
number={C}}

@Chapter{repec:elg:eechap:20188_26,
publisher={Edward Elgar Publishing},
series={Chapters},
edition={None},
booktitle={Handbook of Choice Modelling},
chapter={26},
author={Andrew Daly},
title={Forecasting choice},
year={2024},
month={None},
pages={746-765},
volume={None}
}

@article{HESS2022114800,
title = {The path towards herd immunity: Predicting COVID-19 vaccination uptake through results from a stated choice study across six continents},
journal = {Social Science \& Medicine},
volume = {298},
pages = {114800},
year = {2022},
issn = {0277-9536},
author = {Stephane Hess and Emily Lancsar and Petr Mariel and Jürgen Meyerhoff and Fangqing Song and Eline {van den Broek-Altenburg} and Olufunke A. Alaba and Gloria Amaris and Julián Arellana and Leonardo J. Basso and Jamie Benson and Luis Bravo-Moncayo and Olivier Chanel and Syngjoo Choi and Romain {Crastes dit Sourd} and Helena Bettella Cybis and Zack Dorner and Paolo Falco and Luis Garzón-Pérez and Kathryn Glass and Luis A. Guzman and Zhiran Huang and Elisabeth Huynh and Bongseop Kim and Abisai Konstantinus and Iyaloo Konstantinus and Ana Margarita Larranaga and Alberto Longo and Becky P.Y. Loo and Malte Oehlmann and Vikki O'Neill and Juan {de Dios Ortúzar} and María José Sanz and Olga L. Sarmiento and Hazvinei Tamuka Moyo and Steven Tucker and Yacan Wang and Yu Wang and Edward J.D. Webb and Junyi Zhang and Mark H.P. Zuidgeest}
}

@article{schwarz1978bic,
  title   = {Estimating the Dimension of a Model},
  author  = {Schwarz, Gideon},
  journal = {The Annals of Statistics},
  year    = {1978},
  volume  = {6},
  number  = {2},
  pages   = {461--464},
  doi     = {10.1214/aos/1176344136}
}

@article{akaike1974aic,
  title   = {A New Look at the Statistical Model Identification},
  author  = {Akaike, Hirotugu},
  journal = {IEEE Transactions on Automatic Control},
  year    = {1974},
  volume  = {19},
  number  = {6},
  pages   = {716--723},
  doi     = {10.1109/TAC.1974.1100705}
}

@article{swait1993scale,
  title   = {The Role of the Scale Parameter in the Estimation and Comparison of Multinomial Logit Models},
  author  = {Swait, Joffre and Louviere, Jordan},
  journal = {Journal of Marketing Research},
  year    = {1993},
  volume  = {30},
  number  = {3},
  pages   = {305--314},
  doi     = {10.1177/002224379303000303}
}

@article{brownstone1999forecasting,
  title   = {Forecasting New Product Penetration with Flexible Substitution Patterns},
  author  = {Brownstone, David and Train, Kenneth},
  journal = {Journal of Econometrics},
  year    = {1999},
  volume  = {89},
  number  = {1-2},
  pages   = {109--129},
  doi     = {10.1016/S0304-4076(98)00057-8}
}

@article{SCACCIA202354,
title = {Prediction and confidence intervals of willingness-to-pay for mixed logit models},
journal = {Transportation Research Part B: Methodological},
volume = {167},
pages = {54-78},
year = {2023},
issn = {0191-2615},
author = {Luisa Scaccia and Edoardo Marcucci and Valerio Gatta}
}

@book{mcelreath2020statistical,
  title     = {Statistical Rethinking: A Bayesian Course with Examples in R and Stan},
  author    = {McElreath, Richard},
  year      = {2020},
  edition   = {2},
  publisher = {Chapman and Hall/CRC},
  address   = {Boca Raton, FL},
  isbn      = {9780367139919},
  series    = {Texts in Statistical Science}
}

\end{document}